\documentclass[11pt,a4paper]{article}
    \usepackage{jheppub}
    \usepackage[T1]{fontenc}     
   \usepackage{subcaption}
   \usepackage{float}
   \usepackage{array}
   \usepackage[latin9]{inputenc} 

\DeclareMathOperator{\Tr}{Tr}

\title{Identifying optimal large $N$ limits for marginal $\phi^4$ theory in 4d}

\author{Nadia Flodgren}

\affiliation{The Oscar Klein Centre \& Department of Physics, Stockholm University,\\
AlbaNova, 106 91 Stockholm, Sweden}

\emailAdd{nadia.flodgren@fysik.su.se}

\abstract{We apply our previously developed approach to marginal quartic interactions in multiscalar QFTs, which shows that one-loop RG flows can be described in terms of a commutative algebra, to various models in 4d. We show how the algebra can be used to identify optimal scalings of the couplings for taking large $N$ limits. 
The algebra identifies these limits without diagrammatic or combinatorial analysis. For several models this approach leads to new limits yet to be explored at higher loop orders. We consider the bifundamental and trifundamental models, as well as a matrix-vector model with an adjoint representation. Among the suggested new limit theories are some which appear to be less complex than general planar limits but more complex than ordinary vector models or melonic models.}

\begin{document}

\maketitle

\section{Introduction}\label{sec_intro}

The goal of this paper is to test our algebraic method, developed in \cite{Flodgren:2023lyl} and \cite{Flodgren:2023tri}, on several different models of marginal multiscalar $\phi^4$ theory in 4d and to use it to identify scalings useful for taking large $N$ limits. This results in some new, as well as already known, optimal scalings and limits.

The parameter $N$ measures the size of a global or local symmetry such as $O(N)$ or $U(N)$. There may be other groups and several group size-parameters. When making a $1/N$ expansion and taking the large $N$ limit the objective is finite observables and beta functions, up to a given loop order. In order to achieve this the original couplings $g_k$ of a theory may need to be rescaled w.r.t $N$ such that the rescaled couplings $\lambda_k$ are kept constant as $N \rightarrow \infty$ and $g_k \rightarrow 0$. However, identifying which scaling to use can be non-trivial. 
The scaling of a coupling constant of an interaction is determined by how the field indices contract in the Feynman diagrams. Therefore the methods used to determine scalings often rely on graph theory and combinatorics. 

The concept of optimal scaling was introduced in \cite{Ferrari:2017jgw} for zero dimensional QFTs, where the definition is equivalent to the set of the smallest scaling exponents that give a well-defined large $N$ expansion. The scaling exponent is the power of $N$ that the coupling should be divided by, i.e. $n(k)$ in $\lambda_k=\frac{g_k}{N^{n(k)}}$. 

For matrix models the large $N$ scaling was determined by 't Hooft in \cite{HOOFT1974461}, where he developed a method of obtaining a large $N$ scaling based on graph theory. For these models, the scaling of an operator is according to the number of traces in the interaction. The large $N$ limit is called the planar limit of matrix models because it is dominated by planar diagrams. 
In tensor models, where the rank of the tensors is three or higher, the large $N$ limit is dominated by so called melonic diagrams \cite{Bonzom:2011zz}. 
The optimal scaling for the large $N$ limit of the $O(N)^3$ symmetric tensor model was developed in \cite{Carrozza:2015adg}. 
In the $O(N)^3$ tensor models with quartic interactions the melonic diagram dominance at large $N$ is caused by a particular ``tetrahedral'' coupling which generates the other couplings \cite{Benedetti:2020sye}. The melonic limit can exist in models with tensors of at least rank three, and it is special because a smaller class of planar diagrams dominates, making it much less complicated compared to the planar limit of matrix models. 
Another simpler large $N$ limit is that of vector models, which are know to have a large $N$ limit dominated by a subset of the planar diagrams called bubble diagrams \cite{Coleman:1974jh,Moshe:2003xn}. 

An alternative way to describing large $N$ limit theories in terms of which diagrams dominate is to consider the fields. In this sense limits are vector-, matrix- and tensor-like based on what type of structure the fields span. Vector-like large $N$ limit theories have fields spanned by large vectors with $N$ elements. Similarly, for matrix-like and tensor-like limits the fields span large matrices and tensors respectively. These types of limit theories appear in our analysis and it seems to be the case that they do not always match to the bubble, planar or melon dominated limits. 

Finding alternative limits to the melonic, bubble and planar ones is an active field of research, see for example enhanced tensor models that aim to enhance non-melonic diagrams at large $N$ to investigate different phases of tensor models \cite{Bonzom:2012wa,Bonzom:2015axa,BenGeloun:2017xbd,Geloun:2023oyd}. 
It is also possible for models to have multiple scaling limits, for example when more than one group size-parameter is taken to be large. These types of limits might allow non-melonic diagrams to play a larger role. 
The Veneziano limit of QCD, where the number of colours $N_c$ and flavours $N_f$ are taken to be large while the ratio $N_f/N_c$ is kept fixed, is an example of such a limit \cite{Veneziano:1976wm}. Limits where two parameters are large are sometimes called double-scaling limits, see for example \cite{Ferrari:2017ryl,Azeyanagi:2017mre,Ferrari:2017jgw} where systems of a number $D$ of $N \times N$ matrices are studied in limits where both $N$ and $D$ are large. However, the terminology ``double-scaling limit'' is also applied to limits where one parameter $N$ is large simultaneously as the coupling constant is tuned to a critical value, e.g. \cite{Dartois:2013sra,Bonzom:2022yvc,Bonzom:2021kjy}. There are even triple-scaling limits where there are two large group size-parameters and the coupling constant is tuned to a critical value \cite{Benedetti:2020iyz}. 
For clarity, we choose to avoid the term double-scaling in this work, which does not consider tuning of coupling constants, and instead use the term Veneziano-like for limits where several group-size parameters, like $N$ and $D$, are large. 

In this work we observe that our one-loop algebraic description of RG flow, heretofore applied to multiscalar theories with quartic interactions in 4d \cite{Flodgren:2023lyl,Flodgren:2023tri}, can be used to identify optimal scalings for several matrix and tensor models, and consequently lead to several new, often Veneziano-like, limits for these models. 
For some models a couple different limits result in the same algebras/one-loop beta functions. We consider several models where there are two group size-parameters, similarly to \cite{Ferrari:2017ryl,Azeyanagi:2017mre,Ferrari:2017jgw,Benedetti:2020iyz} but with an approach to identifying the limits that does not rely on graph theory or combinatorics.
  
In section \ref{sec_method} we explain the method of identifying large $N$ limits from one-loop order beta functions. We briefly review the basics of the algebraic description of one-loop order RG flow and then identify the limits via rescalings of the algebra. The possible rescalings lead to a system of linear inequalities of the scaling exponents which forms a linear programming (LP) problem. The redundant conditions of this system, which are normally unimportant in an LP problem, are key to which terms survive for each scaling and limit. This is explained pedagogically using the bifundamental model as an example. 

In section \ref{sec_examples} we demonstrate the capabilities of the method by identifying several limits for two models.
The first model is a matrix-vector version of the trifundamental tensor model, for which we identify four limits that yield distinct algebras, including the previously known vector-like, matrix-like and (melonic) tensor-like limits of the trifundamental model. The fourth limit is a Veneziano-like limit that is matrix-vector-like and results in an algebra less complicated than the matrix-like limit but still more complicated than the tensor-like limit.  
The second model is a system of adjoint scalar multiplets for which we identify five limits that yield distinct algebras, three of which we have described in \cite{Flodgren:2023tri}. Two of the algebras are relatively simple, with one of them sharing several properties with the melonic limit of tensor models. 

In earlier work \cite{Flodgren:2023lyl,Flodgren:2023tri} we used the algebraic concepts of subalgebras and ideals to characterise the RG flow. The focus of this paper is not on this characterisation. Therefore, for the models we consider the subalgebras and ideals for each have been allocated to appendix \ref{AppB}. 

\section{Method}\label{sec_method}

In this section we demonstrate how to calculate the algebra and identify all useful scalings for taking the large $N$ limit. 
We will consider the bifundamental model with symmetry $O(N)_F\times O(M)_F$ as a practical example, because it has several useful scalings and limits.\footnote{The even simpler example of the $O(N)_F$ vector model is mentioned in \cite{Flodgren:2023tri}.}

\subsection{Algebra basics}

The basics of the algebraic description of one-loop RG flow has been described in detail in \cite{Flodgren:2023lyl} and \cite{Flodgren:2023tri}. 
Therefore, what follows is a brief summary. 

The algebraic description arises from the fact that a multiscalar theory in 4d with massless scalars $\phi_A$ and marginal quartic interactions, i.e. with interaction Lagrangian
\begin{equation} 
\begin{split}
\mathcal{L}_{int} = -\frac{1}{4!}\lambda_{ABCD}\phi_A\phi_B\phi_C\phi_D,
\end{split}
\end{equation}
has a leading order RG flow that is quadratic in the quartic interaction tensor $\lambda_{ABCD}$. 
Indices for the irreducible representations of the scalars and any global symmetries are represented by the multi-index $A$.
The quadratic behaviour is seen in the one-loop beta functions
\begin{equation} \label{eq_betaOrg}
\begin{split}
\beta_{ABCD}=\mu\frac{d}{d\mu}\lambda_{ABCD}=\frac{1}{(4\pi)^2}\bigg(\frac{1}{8}\sum_{\text{perms}}\lambda_{ABEF}\lambda_{EFCD} \bigg),
\end{split}
\end{equation}
where the energy scale $\mu$ runs from the IR to the UV \cite{Machacek:1984zw,Luo:2002ti}. The sum is over permutations of the multi-indices and we choose $\lambda_{ABCD}$ to be totally symmetric in its indices. This behaviour allows us to define a product of the couplings
\begin{equation} 
\begin{split}
\lambda \diamond \kappa = \frac{(4\pi)^2}{2}(\beta_{\lambda+\kappa}-\beta_{\lambda}-\beta_{\kappa}). 
\end{split}
\end{equation}
This $\diamond$ product is commutative but not generally associative. The (in general non-associative) algebra is formed by the vector space of the marginal couplings with the $\diamond$ product.

In order to calculate the algebra we introduce a complete basis of symmetric tensor structures $e^k_{ABCD}$ with corresponding coupling constants $\lambda_k$. We assume the basis spans the space of marginal symmetric four-point couplings, labeled by $k$. The marginal quartic interaction is then
\begin{equation} 
\begin{split}
\lambda_{ABCD}=\lambda_ke^k_{ABCD}
\end{split}
\end{equation}
and the one-loop beta functions are
\begin{equation} 
\begin{split}
\beta_{ABCD}=\beta_ke^k_{ABCD},
\end{split}
\end{equation}
where $\beta_{k}=\mu\frac{d\lambda_k}{d\mu}$. From (\ref{eq_betaOrg}) we have that in a multiscalar theory 
\begin{equation} 
\begin{split}
\beta_k=\frac{1}{(4\pi)^2}\lambda_m \lambda_n C^{mn}_k,
\end{split}
\end{equation}
where the coefficients $C^{mn}_k$, which we call structure constants, describe how $\lambda_m$ and $\lambda_n$ appear in the beta function for $\lambda_k$. The structure constants are defined via the product
\begin{equation} \label{eq_prod1}
\begin{split}
(e^m\diamond e^n)_{ABCD} \equiv \frac{1}{8}\sum_{\text{perms}}e^m_{ABEF}e^n_{EFCD}=C^{mn}_ke^k_{ABCD}. 
\end{split}
\end{equation}
The non-associativity of the product occurs because of the symmetrisation via the sum over permutations in (\ref{eq_prod1}). The product is totally symmetric in the multi-indices. For simplicity of notation we will mostly suppress the multi-indices and simply write the product as $e^m\diamond e^n=C^{mn}_ke^k$, where we use Einstein summation convention.  
The algebra will be represented as a symmetric multiplication matrix
\begin{equation}
\left[
\begin{array}{cccc}
e^{1}\diamond e^{1}  & \dots & e^{1}\diamond e^{K}\\
\vdots  & \ddots  & \vdots  \\
e^{K}\diamond e^{1}  &\dots & e^{K}\diamond e^{K}  
\end{array}
\right]
\end{equation}
where each product is expressed using the structure constants.

To demonstrate this algebraic description we will consider the bifundamental model with symmetry $O(N)_F\times O(M)_F$. All the example models we consider have global symmetries that correspond to how the fields transform.\footnote{In \cite{Flodgren:2023lyl} we considered a model with a gauge field but did not describe the gauge field interactions algebraically, even if the corresponding global symmetry was represented.}

\subsubsection{Example: bifundamental model}

The bifundamental model has scalar fields $\phi_A=\phi_{ab}$ where the multi-index $A$ consists of two indices $a$ and $b$, each over a separate orthogonal group. That is, $a=1,\dots,N$ and $b=1,\dots, M$ and the global symmetry is $O(N)_F\times O(M)_F$.\footnote{The symmetry $\mathbb{Z}_2$ that flips the signs of all the fields simultaneously is modded out, meaning the global symmetry is really $O(N)_F\times O(M)_F/\mathbb{Z}_2$.} The fixed points of this model are discussed in for example \cite{RychkovStergiou18}. 

The quartic invariants of the model are the double-trace $\mathcal{O}_1$ and single-trace $\mathcal{O}_2$ interactions
\begin{equation}
\begin{split}
\mathcal{O}_1 &= (\phi_{ab}\phi_{ab})^2 \\
\mathcal{O}_2 &= \phi_{a_1b_1}\phi_{a_1b_2}\phi_{a_2b_1}\phi_{a_2b_2}. 
\end{split}
\end{equation}
The corresponding symmetric tensor structure basis elements are
\begin{equation}
\begin{split}
e^1_{ABCD}=&\delta_{a_1a_2}\delta_{b_1b_2}\delta_{a_3a_4}\delta_{b_3b_4}+\delta_{a_1a_3}\delta_{b_1b_3}\delta_{a_2a_4}\delta_{b_2b_4}+\delta_{a_1a_4}\delta_{b_1b_4}\delta_{a_2a_3}\delta_{b_2b_3} \\
e^2_{ABCD}=&\delta_{a_1a_2}\delta_{b_1b_3}\delta_{a_3a_4}\delta_{b_2b_4}+\delta_{a_1a_2}\delta_{b_2b_3}\delta_{a_3a_4}\delta_{b_1b_4}+\delta_{a_1a_3}\delta_{b_1b_2}\delta_{a_2a_4}\delta_{b_3b_4}\\
+&\delta_{a_1a_3}\delta_{b_1b_4}\delta_{a_2a_4}\delta_{b_2b_3}+\delta_{a_1a_4}\delta_{b_1b_2}\delta_{a_2a_3}\delta_{b_3b_4}+\delta_{a_1a_4}\delta_{b_1b_3}\delta_{a_2a_3}\delta_{b_2b_4}
\end{split}
\end{equation}
which are related to the invariants via
\begin{equation}
\begin{split}
\frac{1}{4!}e^1_{ABCD}\phi_A\phi_B\phi_C\phi_D&=\frac{1}{8}(\phi_{ab}\phi_{ab})^2 \\ 
\frac{1}{4!}e^2_{ABCD}\phi_A\phi_B\phi_C\phi_D&=\frac{1}{4}\phi_{a_1b_1}\phi_{a_1b_2}\phi_{a_2b_1}\phi_{a_2b_2}. \\
\end{split}
\end{equation}
Using this basis we find the finite $N$ and $M$ algebra given by the $2\times 2$ symmetric multiplication matrix 
\begin{equation}\label{eq_NM_alg1}
\left[
\begin{array}{ccc}
e^{1}\diamond e^{1}  & e^2\diamond e^1 \\
e^{2}\diamond e^{1}  &e^2 \diamond e^2
\end{array} 
\right] =
\left[
\begin{array}{ccc}
(8+MN)e^1  & 2(1+M+N)e^1+6e^2 \\
2(1+M+N)e^1+6e^2 & 12e^1+2(4+M+N)e^2
\end{array} 
\right].  
\end{equation}
The algebra (\ref{eq_NM_alg1}) is non-associative, as demonstrated by the fact that $(e^1\diamond e^1)\diamond e^2\neq e^1\diamond(e^1 \diamond e^2)$.
The properties of the algebra in terms of subalgebras and ideals are described in appendix \ref{AppB0}.

In order to easily see how the algebra relates to the beta functions, here are the beta functions for finite $N$ and $M$
\begin{equation} 
\begin{split}
\beta_{\lambda_1}&= \frac{1}{16\pi^2}\big( (8+MN)\lambda_1^2+4(1+M+N)\lambda_1\lambda_2 + 12 \lambda_2^2 \big) \\
\beta_{\lambda_2}&= \frac{1}{16\pi^2}\big( 12\lambda_1\lambda_2 +2(4+M+N)\lambda_2^2 \big). \\
\end{split}
\end{equation}

Next, we are interested in the algebra in the large $N$ and/or $M$ limits. To identify the possible limits that yield a well-defined algebra, the first step is to rescale the coupling constants and basis elements. 

\subsection{General rescaling and parametrisation} \label{sec_parametrisation}
A general rescaling of the coupling constants $\lambda_k \rightarrow \Lambda_k$ and basis elements $e^k\rightarrow E^k$ fulfills 
\begin{equation} 
\begin{split}
\lambda_ke^k&=\Lambda_k E^k. \\
\end{split}
\end{equation}
We will consider models with symmetries containing one or two group size-parameters ($N$ and $M$) with the goal of obtaining finite algebras in the large $N$ and/or $M$ limits. Therefore, we make the scaling ansatz
\begin{equation} \label{eq_scalingdef}
\begin{split}
e^k&=N^{n(k)}M^{m(k)}E^k \\
\lambda_k &= N^{-n(k)}M^{-m(k)}\Lambda_k,
\end{split}
\end{equation}
where the exponents $n(k)$ and $m(k)$ are non-negative real numbers. 
To be able to take a limit with large $N$, large $M$ or both, we choose a parametrisation that defines a relationship between $N$ and $M$
\begin{equation} \label{eq_para}
\begin{split}
M&=v(a)Q^a \\
N&=w(b)Q^b
\end{split}
\end{equation}
where $v(a) \geq 0$ and $w(b)\geq 0$ are constants and we assume the exponents $a\geq 0$ and $b\geq 0$. This parametrisation differs from the one used in \cite{Flodgren:2023tri} and allows more scalings to be identified. 
There are now several possible limits to consider. To take the large $N$ limit with finite $M$ we take $Q$ to be large with $a=0, b>0$. The large $M$ limit with finite $N$ has $Q$ be large with $a>0,b=0$, and limits where $M$ and $N$ are large simultaneously requires $Q$ to be large with $a>0,b>0$. Setting $a=b=0$ leaves us with finite $N$ and $M$. 

In terms of the parametrisation the scalings of the basis elements are
\begin{equation} 
\begin{split}
e^k&=v(a)^{m(k)}w(b)^{n(k)}Q^{u(k)}E^k \\
\end{split}
\end{equation}
where the exponent of $Q$ is 
\begin{equation} 
\begin{split}
u(k)= bn(k)+am(k).
\end{split}
\end{equation}
With this parametrisation $M$ and $N$ are related by
\begin{align} \label{eq_MNk}
M=\frac{v(a)}{w(b)^{a/b}}N^{a/b} 
\end{align}
which shows us that 
\begin{align} \label{eq_c}
c=\frac{a}{b}
\end{align}
is the parameter that determines the balance between $M$ and $N$, which is what separates the limits from each other. 
$c$ is only well defined for $a\geq 0$ and $b>0$, thus $c\geq 0$. However, we can still consider the case $a>0,b=0$ (large $M$, finite $N$) by taking the limit $c \rightarrow \infty$ since if we assume 
 $a>0$
\begin{align} 
\lim_{b \rightarrow 0} \frac{a}{b} = \lim_{b \rightarrow 0} c \rightarrow \infty.
\end{align}
In the rest of the paper we will consider the parametrisation in terms of $c$ and the relation between $M$ and $N$ given by
\begin{align} \label{eq_MNc}
M=\frac{v(bc)}{w(b)^{c}}N^{c}.
\end{align}
To completely get rid of $a$ we define
\begin{equation} \label{eq_u}
\begin{split}
\tilde{u}(k)= \frac{u(k)}{b}=n(k)+cm(k)
\end{split}
\end{equation}
and thus the determining factor for the scaling in the large $Q$ limit is $\tilde{u}(k)$
\begin{equation} \label{eq_rescalingfinal}
\begin{split}
e^k&=v(bc)^{m(k)}w(b)^{n(k)}Q^{b\tilde{u}(k)}E^k. \\
\end{split}
\end{equation}

To be clear, the goal is to identify the scalings, i.e. values of $n(k)$ and $m(k)$ in (\ref{eq_scalingdef}), that allow a finite large $Q$ limit for the algebra/one-loop beta functions. The parametrisation in terms of $c$ (or $a$ and $b$) allows us to classify all possible scalings and limits 
where one or both of $M$ and $N$ are large, and $M$ and $N$ are proportional to each other and related by a positive constant, i.e. $M=pN^c$ where $p\geq 0$ and $c \geq 0$.\footnote{We do not consider negative proportionality constants $p$, which would imply one of $M$ and $N$ to be negative. However, in the context of the group $Sp(N)$ there is an analytic continuation for negative values of $N$, that is $O(-N)\simeq Sp(N)$.}  
We use the practical definition of optimal scaling where it is the set of the smallest scaling exponents, $n(k)$ and $m(k)$ for us, that yield a well-defined large $Q$ expansion \cite{Ferrari:2017jgw}. Lowering a scaling exponent below the optimal value leads to a divergence at large $Q$ while raising one may lead to a fine tuned algebra if less diagrams survive the limit. That is, there are non-optimal but still feasible scalings that yield finite algebras. As we shall see, our method will identify the set of feasible scaling exponents. However, we will focus on the optimal scaling as defined here. In practice this will mean identifying the set of the smallest $\tilde{u}(k)$, defined in (\ref{eq_u}). 

\subsection{Identifying optimal scalings}
Using the general ansatz for our scalings (\ref{eq_rescalingfinal}) we calculate the multiplication matrix for the rescaled basis elements $E^k$. This matrix contains elements with exponents of $Q$ that contain the parameter $c$ and linear combinations of $\tilde{u}(k)$.  
The algebra is well defined in the large $Q$ limit if it has only non-divergent elements in its rescaled multiplication matrix. This is equivalent to all exponents of $Q$ being less than or equal to zero. Applying this requirement to the rescaled algebra we obtain a system of linear inequalities, which we call exponent conditions.

A system of linear inequalities is a linear programming (LP)\footnote{Also called linear optimisation.} problem and can be solved in several ways. 
If the system is small enough it can be solved via Fourier's method\footnote{Also called Fourier-Motzkin elimination.} \cite{Williams86}, and for a system with only two or three variables the problem can be solved graphically. 
For a review of linear programming and a description of the simplex algorithm, which is one method of solving an LP problem, see \cite{GVK180926950}.
Essentially, in an LP problem the inequalities form a convex polytope in the space of variables and this polytope is the region of feasible solutions. In our context the variables are the set of exponents $\tilde{u}(k)$.
The optimal solution to an LP problem exists on the boundary of the polytope.\footnote{In an LP problem there is a linear objective function that one wants to minimise (or maximise) on the polytope. The optimal solution is the minimum (or maximum) value of the objective function, which is always found on the boundary of the polytope.} 
For the systems considered in this paper Fourier's method is enough, although we use the simplex algorithm via Mathematica.  

We use the word optimal in two ways, the \textit{optimal solution} refers to the optimal solution to an LP problem (in our case identified via the simplex algorithm) and \textit{optimal scaling} refers to the scaling that is the set of the smallest scaling exponents that give a well-defined large $N$ expansion. The optimal solution we find based on the exponent conditions is what will correspond to the optimal scaling.

Note that some of the inequalities in an LP problem may be redundant, i.e. implied by other inequalities. For example, $x \geq 0$ is redundant if we also have $x \geq 1$. 
The redundant conditions can be removed from the system of inequalities without any effect on the solutions. Non-redundant conditions are called essential.
However, in our context redundant conditions are very important because they always correspond to vanishing terms of the algebra in the large $Q$ limit. 
Terms that survive the limit correspond to essential conditions being fulfilled as equalities. 
If an essential condition is fulfilled by an inequality then the corresponding terms in the algebra vanish in the large $Q$ limit.

For example, the essential condition $x \geq 1$ corresponds to terms with factors $Q^{-x+1}$. When $x=1$, i.e. the essential condition is fulfilled by an equality, the terms survive the large $Q$ limit, but when $x>1$ the terms vanish in the same limit. 
The redundant condition $x \geq 0$ corresponds to terms with factors $Q^{-x}$ which always vanish at large $Q$ because of the essential condition $x\geq 1$.
If the goal is to preserve as many terms as possible in the large $Q$ limit then we want to fulfill as many essential exponent conditions as equalities as possible.\footnote{Note that the same exponent condition can appear multiple times (some will always appear multiple times because the multiplication matrix is symmetric), meaning that multiple terms of the algebra will vanish or not depending on how this conditions is fulfilled.} 

It is important to note that our system of inequalities contains the free parameter $c$. Since $c$ is not determined there may be multiple optimal solutions to the LP problem, each valid for a different value of $c$. We shall see that this is indeed the case for the bifundamental model. 

The goal of our approach is to obtain a finite description of the RG flow at large $N$. However, one might want to obtain other finite quantities or observables in the large $N$ limit. In appendix A.5 of \cite{Jepsen:2023pzm} a similar method to ours is used to obtain several (a line segment of) optimal scalings in the large $N$ limit. The difference compared to our method is that they obtain the scaling exponent conditions from the free energy density instead of the beta functions via the algebra.

\subsubsection{Example: bifundamental model}
For the bifundamental model the list of exponent conditions is (\ref{eq_ineq0})-(\ref{eq_ineq1})
\begin{equation} \label{eq_ineq0}
\begin{split}
\tilde{u}(1) &\geq 0 \\
\tilde{u}(2) &\geq 0 \\
\end{split}
\end{equation}
\begin{equation} \label{eq_ineq1}
\begin{split}
\tilde{u}(1) &\geq 1+c \\
\tilde{u}(2) &\geq 1 \\
\tilde{u}(2) &\geq c \\
\tilde{u}(2) &\geq \frac{\tilde{u}(1)}{2}.
\end{split}
\end{equation}
The conditions (\ref{eq_ineq0}) are redundant and therefore correspond to vanishing terms of the algebra in the large $Q$ limit.%
\footnote{As part of our method we assume that $\tilde{u}(k)\geq 0$ but they also appear as redundant exponent conditions (\ref{eq_ineq0}) for this model. As redundant exponent conditions they correspond to vanishing terms in the algebra at large $Q$, but note that if they had only appeared as assumptions and not redundant exponent conditions, they would not correspond to vanishing terms in the limit.}
Whether conditions containing $c$ are redundant or essential depends on the value of $c$. Therefore, not all conditions (\ref{eq_ineq1}) are essential for all values of $c$. 
The two-dimensionality of the system of inequalities means we can solve it graphically. 
From the exponent conditions (\ref{eq_ineq1}) we note that for $c=1$ two of the conditions become equivalent. This affects the solution to the system because it implies that there are two different sets of essential inequalities, one for $c\leq1$ and one for $c\geq 1$, that are identical for $c=1$.
Each set therefore leads to a different optimal solution to the LP problem, which in turn corresponds to a different line segment of optimal scalings parametrised by $c$. 

Table \ref{TONM_exp1} shows which of the four exponent conditions in (\ref{eq_ineq1}) are fulfilled by equalities or strict inequalities for each optimal solution and its range. 
The table shows that most exponent conditions are fulfilled by an equality where the two solutions overlap at $c=1$. For $0 \leq c<1$ and $c>1$ different sets of exponent conditions are fulfilled by equalities, which corresponds to the two different optimal solutions, one for each range of $c$. 
 \begin{table}[h]
\centering
\begin{tabular}{|c|c|c|c|}
  \hline
Exp. cond. & $0 \leq c<1$ & $c=1$ & $c>1$  \\ \hline\hline
$\tilde{u}(1)\geq1+c$  & = &  =  & =  \\ \hline
$\tilde{u}(2)\geq 1$  & = & =   & >  \\ \hline
$\tilde{u}(2)\geq c$  & > & =  & =  \\ \hline
$\tilde{u}(2)\geq \frac{\tilde{u}(1)}{2}$  & > & = & > \\ \hline
\end{tabular}
\caption{Exponent conditions fulfilled by an equality or strict inequality for the optimal solutions to the LP problem (\ref{eq_ineq1}) and their ranges in $c$. Recall $M \propto N^c$.}
\label{TONM_exp1}
\end{table}

For $0 \leq c\leq 1$ the optimal solution to (\ref{eq_ineq1}) is $\tilde{u}(1)=1+c$ and $\tilde{u}(2)=1$ which determines $n(k)$ and $m(k)$ in (\ref{eq_scalingdef}) via (\ref{eq_u}). This corresponds to the scaling 
\begin{align} \label{eq_NM_scaling1}
\lambda_1=\frac{\Lambda_1}{MN} && \lambda_2=\frac{\Lambda_2}{N}
\end{align}
which yields a finite beta function in the large $Q$ limit for $0 \leq c\leq 1$ and where $M$ and $N$ are related by (\ref{eq_MNc}).
In the range $c \geq 1$ the optimal solution to (\ref{eq_ineq1}) yields 
\begin{align} \label{eq_NM_scaling2}
\lambda_1=\frac{\Lambda_1}{MN} && \lambda_2=\frac{\Lambda_2}{M},
\end{align}
again where $M$ and $N$ are related by (\ref{eq_MNc}). 
Note that whenever two optimal scalings overlap for a value of $c$ the two scalings will be equivalent with respect to the large group size-parameters and yield equivalent algebras, i.e. algebras that are identical under a rescaling of the basis elements and couplings by a constant.

For each range of $c$ the scaling exponents $n(k)$ and $m(k)$ are determined but to find which scaling is optimal we need to specify $c$ (the relation between $M$ and $N$) completely. 
For example, using (\ref{eq_MNc}) and setting $c=0$ means $M=v(0)$ is a constant, while setting $c=1$ means $M=\frac{v(b)}{w(b)}N$. Clearly, inserting $M=v(0)$ and $M=\frac{v(b)}{w(b)}N$ into (\ref{eq_NM_scaling1}) yields two different optimal scalings.
However, to keep the notation simple we will refer to the scalings (\ref{eq_NM_scaling1}) and  (\ref{eq_NM_scaling2}) as optimal for their respective ranges of $c$. 

Note that for each scaling the limits $N \rightarrow \infty$ and $M \rightarrow \infty$ do not commute for all relations between $M$ and $N$. For example, scaling (\ref{eq_NM_scaling1}) is only optimal for $0\leq c \leq 1$ which corresponds to $N \geq M$. We cannot use scaling (\ref{eq_NM_scaling1}) and take the limit where $N$ is finite and $M$ large because then $N \geq M$ is not fulfilled and the resulting algebra will have divergences. Similarly, for scaling (\ref{eq_NM_scaling2}) we cannot take the limit where $M$ is finite and $N$ is large without obtaining divergences in the algebra. 

\begin{figure}
     \centering
     \begin{subfigure}[b]{0.3\textwidth}
         \centering
         \includegraphics[width=\textwidth]{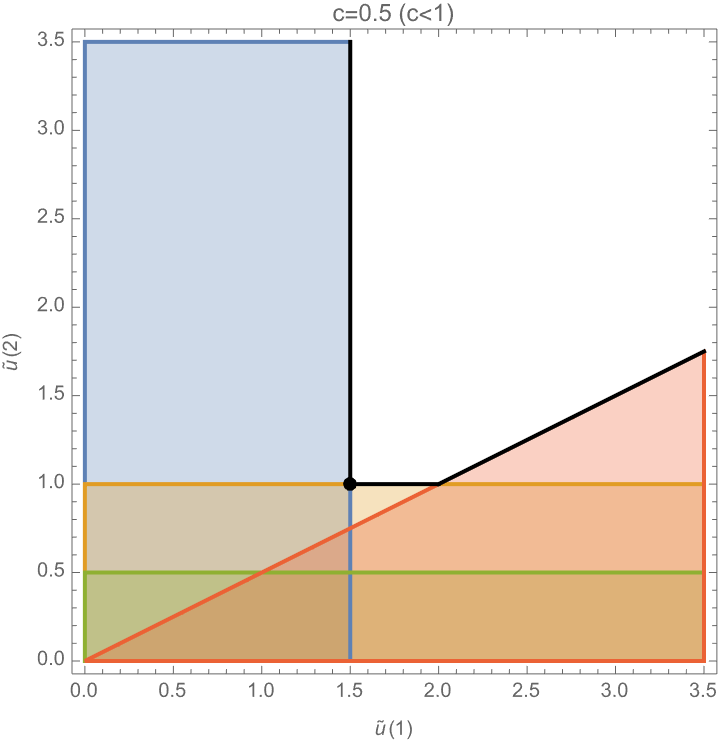}
         \caption{$c=0.5$ $(c<1)$}
         \label{fig:c05}
     \end{subfigure}
     \hfill
     \begin{subfigure}[b]{0.3\textwidth}
         \centering
         \includegraphics[width=\textwidth]{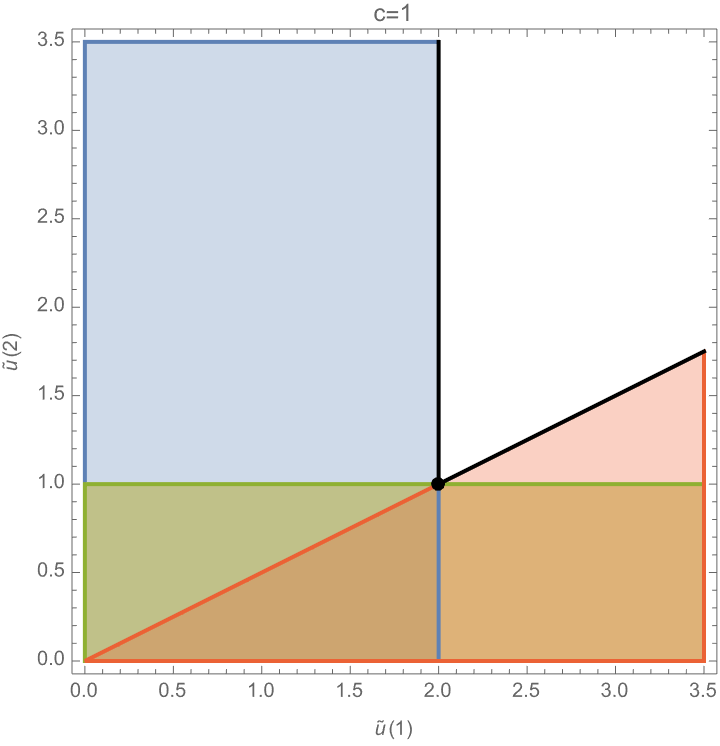}
         \caption{$c=1$}
         \label{fig:c1}
     \end{subfigure}
     \hfill
    \begin{subfigure}[b]{0.3\textwidth}
         \centering
         \includegraphics[width=\textwidth]{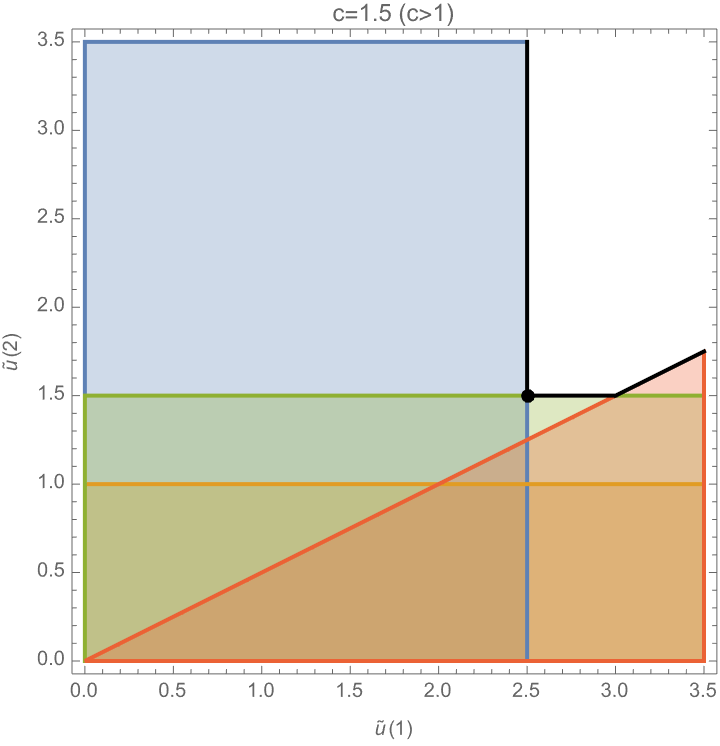}
         \caption{$c=1.5$ $(c>1)$}
         \label{fig:c15}
     \end{subfigure}
     \hfill
        \caption{Exclusion plot showing the feasible region of solutions (white with black edge) to the constraints (\ref{eq_ineq1}) for $0\leq c<1$, $c=1$ and $c>1$. The optimal solutions for specific values of $c$ are shown as black dots.}
        \label{fig:ONM}
\end{figure}

Figure \ref{fig:ONM} 
shows the feasible region of solutions and the optimal solution to (\ref{eq_ineq1}) for example values of $c$ on the two line segments $0\leq c<1$, $c>1$ and where they overlap at $c=1$.
Note that the feasible region is an unbounded polygon.

Recall from section \ref{sec_parametrisation} that the optimal scaling is the set of the smallest $\tilde{u}(k)$ for a given value of $c$. 
However, all feasible values of $\tilde{u}(k)$ yield well-defined algebras in the large $Q$ limit, but most of these will be trivial, i.e. all terms of the algebra will vanish. 
In the interior of the feasible region (the white area in figure \ref{fig:ONM}) all relevant exponent conditions (\ref{eq_ineq1}) are fulfilled by inequalities, which we recall corresponds to vanishing terms in the algebra at large $Q$. Consequently, all scalings in the interior of the feasible region only give vanishing algebras. For the solutions on the edges of the feasible region at least one exponent condition is fulfilled by an equality, meaning that at least one term survives the limit. At vertices on the edges at least two exponent conditions are fulfilled by equalities, meaning at least two terms survive. The optimal solution is the vertex where the set of $\tilde{u}(k)$ have the smallest values.

This does not mean that non-optimal solutions are useless. In figures \ref{fig:c05} and \ref{fig:c15} the non-optimal solution that is the vertex to the right of the optimal one is a feasible solution and will yield a fined tuned algebra. These non-optimal solutions and the scalings they yield are analysed in greater detail in the appendix \ref{AppC}. 
Understanding which diagrams survive for each scaling and limit might be interesting in the context of enhancing certain non-dominant types of diagrams.

\subsection{Limits and algebras}

In this section we discuss the non-commutativity of the limits obtained via the optimal scalings, and the two different types of limits the method yields.  

We shall see that for all models we consider with symmetries containing two group size-parameters $N$ and $M$, we can identify two scalings, each optimal for a range of values of $c$ (recall $M\propto N^c$), just as for the bifundamental model, which we can see as two line segments of optimal scalings parametrised by $c$. 
Because of this the large $N$ and $M$ limits will in general not commute. 
 
To generalise the example of the bifundamental model, consider a model with two scalings that overlap for $c=c_*>0$, i.e. scaling 1 is optimal for $0\leq c \leq c_*$ and scaling 2 for $c\geq c_*$. 
This implies that scaling 1 is optimal for $N^{c_*}\geq M$ and scaling 2 is optimal for $M \geq N^{c_*}$. In general, for each scaling the limits $N\rightarrow \infty$ and $M \rightarrow \infty$ do not commute. For scaling 1 we cannot take $N$ to be finite and $M \rightarrow \infty$ because then $N^{c_*}\geq M$ no longer holds and the limit will yield an algebra with divergent terms. The same is true for scaling 2, for which we cannot take the large $N$ finite $M$ limit. 
For Veneziano-like limits both $N$ and $M$ are large but the relation between them is fixed, e.g. $M/N^c$ is fixed for some $c$, meaning that we are not taking two separate limits as much as one limit, e.g. by taking $N$ to be large after expressing $M$ in terms of $N$ via $M\propto N^c$.

The non-commutativity of the limits has been observed before. In \cite{Ferrari:2017jgw} matrix-tensor models of symmetry $O(N)^2 \times O(D)^r$ are shown to have an optimal scaling for which the limits $N\rightarrow \infty$ and $D \rightarrow \infty$ do not commute. The optimal scaling will yield divergences in the limit where $N$ is finite and $D$ large, which means $N \rightarrow \infty$ must be taken first. 
What we add here is that for the models we consider there is a second optimal scaling for which $M \rightarrow \infty$ must be taken before $N \rightarrow \infty$.

With our line segments of optimal scalings parametrised by $c$ identified, we can take all possible large $M$ and/or $N$ limits (via large $Q$) and find out what algebras they yield. 
There are two types of limits (here we have specified to optimal scalings but non-optimal scalings yield the same types of limits, see appendix \ref{AppC}).
\begin{itemize}
\item \textbf{Critical limit:} a large $Q$ limit using optimal scaling at specific values of $c$ where two or more exponent conditions, one of which has to be essential, become identical. The overlap of essential conditions means there are fewer redundant conditions and therefore more terms of the algebra that survive the limit, compared to the reduced limits. 
\item \textbf{Reduced limit:} a large $Q$ limits using optimal scaling for values of $c$ where no essential exponent condition overlaps with another exponent condition, i.e. the opposite of a critical limit. These occur for ranges of $c$ since overlaps happen for specific values of $c$. Compared to the critical limits fewer terms of the algebra survive because there are more redundant exponent conditions for these values of $c$.    
\end{itemize}
Reduced limits by definition have fewer terms than the critical limits that are adjacent on the spectrum of $c$. They also often have fewer free parameters. In the models we consider, which all have two group size-parameters ($N$ and $M$), the critical limits all yield algebras that have one free parameter while the reduced limits algebras have none, with our optimal scalings. 

Note that critical limits may occur for values of $c$ where two optimal scalings overlap, but also at $c=0$ or $c \rightarrow \infty$ which are at the ends of the spectrum of $c$ and therefore do not correspond to overlaps of scalings. At values of $c$ where two optimal scalings overlap the critical limit may be taken using either of the two scalings because they are equivalent w.r.t the large group size-parameters. The two versions of the limit will yield algebras that are equivalent, in the sense that they are directly related by a rescaling of the elements w.r.t a constant. 
 
The algebras obtained in the different limits of one model can be identical. 
The symmetry of the model, for example, may yield the same algebra in several limits. This will become clear in the example of the bifundamental model where the symmetry $O(N)_F\times O(M)_F$ treats $N$ and $M$ identically, which implies that the algebras in the limits where only one of $N$ and $M$ is large and the other is finite should be identical. 

\subsubsection{Example: bifundamental model} \label{sec_ONM_final}
\begin{figure}[!htb] 
\begin{center}
\includegraphics[scale=0.5]{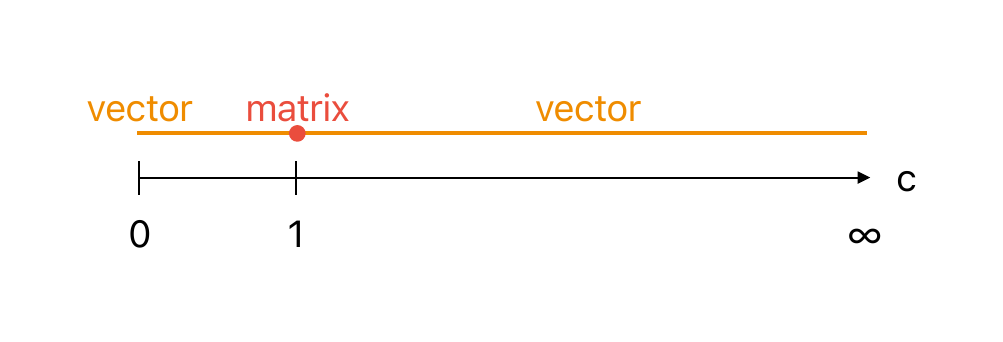}
\caption{The distribution of the algebras on the spectrum of $c$ for the $O(N)\times O(M)$ model. Dots represent critical algebras. Lines represent reduced algebras. The same color represents identical algebras. Recall $M\propto N^c$. The labels show which type of limit yields the algebra.} 
\label{Fig:CspecNM}
\end{center}
\end{figure}

For the bifundamental model there is only one overlap of exponent conditions which occurs for $c=1$ and corresponds to an overlap of two optimal scalings. This means that there is one critical limit for $c=1$ and two reduced limits for $0 \leq c <1$ and $c>1$. 
We identified two scalings (\ref{eq_NM_scaling1}) and (\ref{eq_NM_scaling2}), optimal for the ranges $0 \leq c \leq 1$  and $c\geq 1$ (including $c \rightarrow \infty$) respectively. 
Due to the symmetry of the model we expect (and find) that the two reduced limits yield identical algebras. The limits and resulting algebras are detailed below. 
\begin{itemize}
\item \textbf{Reduced limit $0\leq c <1$:} a vector-like large $N$ and $M$ limit where $M \propto N^c$ for $0\leq c<1$, i.e. $N \gg M$. The algebra is (\ref{eq_ONM_mat1}).
\item \textbf{Critical limit $c=1$:} a matrix-like large $N$ and $M$ limit where $M/N=\omega$ is fixed. $\omega \geq 0$ is called a scaling coefficient. The algebra is (\ref{eq_ONM_mat2}), or equivalently (\ref{eq_ONM_mat3}). 
Taking $\omega=0$ in (\ref{eq_ONM_mat2}), or $\omega \rightarrow \infty$ in (\ref{eq_ONM_mat3}), makes some structure constants of the algebra vanish and yields the reduced algebra (\ref{eq_ONM_mat1}).
\item \textbf{Reduced limit $c>1$:} a vector-like large $N$ and $M$ limit where $M \propto N^c$ for $c>1$, i.e. $M \gg N$. 
The algebra is (\ref{eq_ONM_mat1}). 
\end{itemize}
Below we discuss the properties of the algebras. Figure \ref{Fig:CspecNM} shows how the algebras are distributed on the spectrum of $c$.

The critical limit is a Veneziano-like limit obtained for $c=1$, i.e. when both $M$ and $N$ are large and $M/N=\omega$ is fixed. The scaling coefficient is defined as $\omega \equiv \frac{v(b)}{w(b)}\geq 0$ and comes from (\ref{eq_MNc}) for $c=1$. This limit is matrix-like in the sense that $\phi_{ab}$ now spans a large $N\times \omega N$ matrix. 

This critical limit occurs when the two optimal scalings (\ref{eq_NM_scaling1}) and (\ref{eq_NM_scaling2}) are equivalent w.r.t the large group size-parameters. We can observe this equivalence by replacing $M$ (or $N$) by $M=\omega N$ in each scaling 
\begin{align}\label{eq_NM_opt1}
\lambda_1=\frac{\Lambda_1}{\omega N^2} && \lambda_2=\frac{\Lambda_2}{N}.
\end{align}
\begin{align}\label{eq_NM_opt2}
\lambda_1=\frac{\Lambda_1}{\omega N^2} && \lambda_2=\frac{\Lambda_2}{\omega N}.
\end{align}
The only difference between (\ref{eq_NM_opt1}) and (\ref{eq_NM_opt2}) is that $\lambda_2$ scales differently with respect to the constant $\omega$, which does not affect the properties of the algebra in the limit. Using the optimal scaling (\ref{eq_NM_opt1}), taking $N$ to be large and discarding subleading terms the resulting algebra is
\begin{equation}\label{eq_ONM_mat2}
\begin{bmatrix}
E^1 & 2(1+\omega)E^1 \\
2(1+\omega)E^1 & 12\omega E^1+2(1+\omega)E^2 \\
\end{bmatrix}
\end{equation}
with the free parameter $\omega$. 
Taking the large $N$ limit but using the optimal scaling (\ref{eq_NM_opt2}) instead yields an equivalent algebra (\ref{eq_ONM_mat3}) with the free parameter $\frac{1}{\omega}$. Replacing $\omega \rightarrow \frac{1}{\omega}$ in (\ref{eq_ONM_mat2}) gives the algebra (\ref{eq_ONM_mat3}). 
\begin{equation}\label{eq_ONM_mat3}
\begin{bmatrix}
E^1 & 2(1+\frac{1}{\omega})E^1 \\
2(1+\frac{1}{\omega})E^1 & \frac{12}{\omega}E^1+2(1+\frac{1}{\omega})E^2 \\
\end{bmatrix}.
\end{equation}
These equivalent algebras are non-associative, see e.g. $E^1\diamond (E^2 \diamond E^2) \neq (E^1 \diamond E^2)\diamond E^2$.

To compare to known results for the $O(N)^2$ model we set $\omega=1$ (i.e. $M=N$), which reduces the symmetry to $O(N)^2$, although no terms vanish in the algebra. The scalings become the optimal scaling used in for example \cite{Giombi:2017dtl,Jepsen:2023pzm}, and the algebra (\ref{eq_ONM_mat2}) becomes the $O(N)^2$ model algebra/one-loop beta functions. 

The reduced limits with $M\propto N^c$ where $0\leq c< 1$ and $c>1$ include the vector-like limits where only one of $M$ and $N$ is large via $c=0$ and $c \rightarrow \infty$. The limits with $0<c<1$ and $1<c<\infty$ are almost vector-like in the sense that one of $M$ and $N$ is much larger than the other, e.g. for $c>1$ we have $M \gg N$. 
After discarding subleading terms these limits all yield the same algebra
\begin{equation} \label{eq_ONM_mat1}
\begin{bmatrix}
E^1 & 2E^1 \\
2E^1 & 2E^2 \\
\end{bmatrix}.
\end{equation}
Note that no free parameter appears in this algebra and that it is a special case of the critical algebra (\ref{eq_ONM_mat2}) for $\omega=0$, and the equivalent algebra (\ref{eq_ONM_mat3}) for $\omega \rightarrow \infty$. This algebra is associative.

\section{Matrix-vector models} \label{sec_examples}

To demonstrate the power of the method, we show the optimal scalings and limit algebras for two models. First, we consider a matrix-vector version of the trifundamental tensor model and identify four different limits yielding distinct algebras, three of which have been analysed before in \cite{Benedetti:2020sye}. Then we return to a model of $M$ scalar multiplets in the adjoint representation of $SU(N)$ which we studied in \cite{Flodgren:2023lyl,Flodgren:2023tri}. In \cite{Flodgren:2023tri} we identified three limits and here we complement that analysis with two new limits.  

\subsection{$O(N)_F^2 \times O(M)_F$} \label{sec_ONNM}

The trifundamental tensor model has tensor fields $\phi_{A}$, where the multi-index $A=abc$ and each index $a,b,c$ belongs to an $O(N)$ group. The  global symmetry is $O(N_1)\times O(N_2) \times O(N_3)$ and the three positive integers $N_1, N_2$ and $N_3$ can all be taken to be large.\footnote{The symmetry $\mathbb{Z}_2^3$ needs to be modded out, but this does not affect the following analysis, similarly to in \cite{Benedetti:2020sye}.} We will consider a slightly less general version with the symmetry $O(N)^2 \times O(M)$ and take the limits where $N$ or $M$ or both are large. 
As we shall see, our method allows us to identify several limits, such as a vector-like limit, a matrix-like limit, a tensor-like limit that includes the melonic limit of $O(N)^3$, as well as a matrix-vector-like limit where $M/N^2$ is fixed. 
All of these limits except the matrix-vector-like one have been analysed in \cite{Benedetti:2020sye} in $d=4-\epsilon$ dimensions up to two-loop order. The matrix-vector-like limit has similarities to limits considered in \cite{Ferrari:2017jgw,Benedetti:2020iyz}.

The trifundamental model has five quartic invariant interactions, named after their shapes they are a double-trace one $\mathcal{O}_1$, three pillows $\mathcal{O}_2,\mathcal{O}_3,\mathcal{O}_4$ and a tetrahedron $\mathcal{O}_5$, see \cite{Benedetti:2020sye} for diagrammatic representations. 
The explicit expressions for the invariants and the corresponding symmetric basis elements are given in appendix \ref{AppA1}. 
The symmetric basis elements are related to the invariants via
\begin{equation} \label{eq_ONNM_brel}
\begin{split}
\frac{1}{4!}e^1_{ABCD}\phi_A\phi_B\phi_C\phi_D &= \frac{1}{8}(\phi_{a b c} \phi_{a b c})^2  \\
\frac{1}{4!}e^2_{ABCD}\phi_A\phi_B\phi_C\phi_D &= \frac{1}{4}\phi_{a_1 b_1 c_1} \phi_{a_1 b_1 c_2} \phi_{a_2 b_2 c_1} \phi_{a_2 b_2 c_2}  \\
\frac{1}{4!}e^3_{ABCD}\phi_A\phi_B\phi_C\phi_D &= \frac{1}{4}\phi_{a_1 b_1 c_1} \phi_{a_1 b_2 c_1} \phi_{a_2 b_1 c_2} \phi_{a_2 b_2 c_2}  \\
\frac{1}{4!}e^4_{ABCD}\phi_A\phi_B\phi_C\phi_D &= \frac{1}{4}\phi_{a_1 b_1 c_1} \phi_{a_2 b_1 c_1} \phi_{a_1 b_2 c_2} \phi_{a_2 b_2 c_2}  \\
\frac{1}{4!}e^5_{ABCD}\phi_A\phi_B\phi_C\phi_D &= \frac{1}{4}\phi_{a_1 b_1 c_1} \phi_{a_1 b_2 c_2} \phi_{a_2 b_1 c_2} \phi_{a_2 b_2 c_1}. \\
\end{split}
\end{equation}

\subsubsection{Limits and algebras}\label{sec_ONMM_a}

By applying the method described in section \ref{sec_method}, i.e. rescaling the couplings in order to identify well-defined algebras in the limits, we obtain a set of exponent conditions from which we find two line segments of optimal scalings parametrised by $c$ that 
overlap for $c=2$ (recall that $M \propto N^c$). 
Scaling (\ref{eq_NMM_scaling1}) is optimal for $0\leq c\leq 2$, i.e. for $N^2 \geq M$,
\begin{align} \label{eq_NMM_scaling1}
\lambda_1=\frac{\Lambda_1}{MN^2} && \lambda_2=\frac{\Lambda_2}{MN} && \lambda_3=\frac{\Lambda_3}{MN} && \lambda_4=\frac{\Lambda_4}{N^2} &&  \lambda_5=\frac{\Lambda_5}{N\sqrt{M}}.
\end{align}
Scaling (\ref{eq_NMM_scaling2}) is optimal for $c\geq 2$, i.e. for $M \geq N^2$,
\begin{align} \label{eq_NMM_scaling2}
\lambda_1=\frac{\Lambda_1}{MN^2} && \lambda_2=\frac{\Lambda_2}{MN} && \lambda_3=\frac{\Lambda_3}{MN} && \lambda_4=\frac{\Lambda_4}{M} &&  \lambda_5=\frac{\Lambda_5}{M}.
\end{align}
Note that in both scalings the first three coupling constants scale the same way, indicating that the behaviour of the pillow $E^4$ and tetrahedron $E^5$ differs the most between the limits. 
For each scaling the large $N$ and $M$ limits do not commute, as we cannot take $N$ to be finite for (\ref{eq_NMM_scaling1}) or $M$ finite for (\ref{eq_NMM_scaling2}), while $M$ or $N$, respectively, is large.

In total there are four large $M$ and/or $N$ limits that yield distinct algebras. The scalings overlap at $c=2$ but the exponent conditions show that both $c=2$ and $c=0$ indicate critical limits, while the rest of the values of $c$ correspond to reduced limits. Similarly to the bifundamental model, the overlap of the scalings at $c=2$ means that the algebras in this limit are equivalent up to a rescaling of the couplings.
The limits are listed below.
\begin{itemize}
\item \textbf{Critical limit $c=0$:} 
a matrix-like large $N$ limit where $M$ is a constant free parameter. The algebra is (\ref{eq_NMM_A1})-(\ref{eq_NMM_A2}) with (\ref{eq_NMM_A1}) and (\ref{eq_NMM_A2}) showing columns 1-3 and 4-5 of the $5 \times 5$ symmetric matrix, respectively.
Taking the limit $M \rightarrow \infty$ in the resulting algebra takes some structure constants to zero and produces the reduced limit algebra with $0<c<2$ (\ref{eq_NMM_B}).
\item \textbf{Reduced limit $0<c<2$:} a tensor-like large $N$ and $M$ limit where $M\propto N^c$ for $0<c<2$, i.e. $N^2 \gg M$. The algebra is (\ref{eq_NMM_B}). This limit includes the melonic limit ($c=1$) and is in this sense tensor-like.
\item \textbf{Critical limit $c=2$:} a matrix-vector-like large $N$ and $M$ limit where $M/N^2=\rho^2$ is fixed. $\rho\geq 0$ is a scaling coefficient defined as $\rho \equiv +\frac{\sqrt{v(2b)}}{w(b)}$ based on the relation (\ref{eq_MNc}) for $c=2$. 
Scalings  (\ref{eq_NMM_scaling1}) and (\ref{eq_NMM_scaling2}), respectively yield the equivalent algebras (\ref{eq_NMM_C}) and (\ref{eq_NMM_D}) in this limit. Setting $\rho=0$ in (\ref{eq_NMM_C}) gives the reduced limit algebra with $0<c<2$ (\ref{eq_NMM_B}), and taking $\rho \rightarrow \infty$ in (\ref{eq_NMM_D}) gives the reduced limit algebra with $c>2$ (\ref{eq_NMM_EF}).
\item \textbf{Reduced limit $c>2$:} a vector-like large $M$ and $N$ limit where $M \propto N^c$ for $c>2$, i.e. $M \gg N^2$. 
The algebra is (\ref{eq_NMM_EF}). 
\end{itemize}
The algebras are given by the multiplication tables (\ref{eq_NMM_A1})-(\ref{eq_NMM_EF}), where $tri$ labels the model $O(N)_F^2\times O(M)_F$, since it is a matrix-vector version of the trifundamental model. The $C$ and $R$ stand for critical and reduced respectively. The lower index number indicates the range/value of $c$. Equivalent algebras have matrices distinguished by an $a$ and $b$. 
\begin{equation} \label{eq_NMM_A1} 
C_{0(1-3)}^{tri}=2\begin{bmatrix}
\frac{E^1}{2} & (1+\frac{1}{M})E^1 & (1+\frac{1}{M})E^1 \\ 
(1+\frac{1}{M})E^1 & \frac{6}{M}E^1+(1+\frac{1}{M})E^2 & 2E^1+\frac{1}{M}E^2+\frac{1}{M}E^3+\frac{2}{M^2}E^4 \\
(1+\frac{1}{M})E^1 & 2E^1+\frac{1}{M}E^2+\frac{1}{M}E^3+\frac{2}{M^2}E^4 & \frac{6}{M}E^1+(1+\frac{1}{M})E^3 \\
E^1 & 2E^1+\frac{1}{M}E^4 & 2E^1+\frac{1}{M}E^4 \\
\frac{2}{\sqrt{M}}E^1 & \frac{2}{\sqrt{M}}E^1+\frac{1}{\sqrt{M}}E^2+\frac{2}{M^{3/2}}E^4+\frac{1}{M}E^5 & \frac{2}{\sqrt{M}}E^1+\frac{1}{\sqrt{M}}E^3+\frac{2}{M^{3/2}}E^4+\frac{1}{M}E^5
\end{bmatrix}
\end{equation}
\begin{equation} \label{eq_NMM_A2}
C_{0(4-5)}^{tri}=2\begin{bmatrix}
E^1 & \frac{2}{\sqrt{M}}E^1 \\
2E^1+\frac{1}{M}E^4 & \frac{2}{\sqrt{M}}E^1 + \frac{1}{\sqrt{M}}E^2+\frac{2}{M^{3/2}}E^4+\frac{1}{M}E^5 \\
2E^1+\frac{1}{M}E^4 & \frac{2}{\sqrt{M}}E^1+\frac{1}{\sqrt{M}}E^3+\frac{2}{M^{3/2}}E^4+\frac{1}{M}E^5 \\
E^4 & \frac{2}{\sqrt{M}}E^4 \\
\frac{2}{\sqrt{M}}E^4 & E^2+E^3+(1+\frac{2}{M})E^4
\end{bmatrix}
\end{equation}
\begin{equation}\label{eq_NMM_B}
R_{0<c<2}^{tri}=2\begin{bmatrix}
\frac{E^1}{2} & E^1 & E^1 & E^1 & 0 \\
E^1 & E^2 & 2E^1 & 2E^1 & 0 \\
E^1 & 2E^1 & E^3 & 2E^1 & 0 \\
E^1 & 2E^1 & 2E^1 & E^4 & 0 \\
0 & 0 & 0 & 0 & E^2+E^3+E^4 \\
\end{bmatrix}
\end{equation}
\begin{equation} \label{eq_NMM_C}
C_{2a}^{tri}=2\begin{bmatrix}
\frac{E^1}{2} & E^1 & E^1 & (1+\rho^2)E^1 & \rho E^1 \\
E^1 & E^2 & 2E^1 & 2E^1+\rho^2E^2 & \rho E^2 \\
E^1 & 2E^1 & E^3 & 2E^1+\rho^2E^3 & \rho E^3 \\
(1+\rho^2)E^1 & 2E^1+\rho^2E^2 & 2E^1+\rho^2E^3 & 6\rho^2E^1+(1+\rho^2)E^4 & 2\rho E^1+\rho^2E^5 \\
\rho E^1 & \rho E^2 & \rho E^3 & 2\rho E^1+\rho^2E^5 & E^2+E^3+E^4 \\
\end{bmatrix}
\end{equation}
\begin{equation} \label{eq_NMM_D}
C_{2b}^{tri}=
2\begin{bmatrix}
\frac{E^1}{2} & E^1 & E^1 & (1+\frac{1}{\rho^2})E^1 & E^1 \\
E^1 & E^2 & 2E^1 & \frac{2}{\rho^2}E^1+E^2 & E^2 \\
E^1 & 2E^1 & E^3 & \frac{2}{\rho^2}E^1+E^3 & E^3 \\
(1+\frac{1}{\rho^2})E^1 & \frac{2}{\rho^2}E^1+E^2 & \frac{2}{\rho^2}E^1+E^3 &\frac{6}{\rho^2}E^1+(1+\frac{1}{\rho^2})E^4 & \frac{2}{\rho^2}E^1+E^5 \\
E^1 & E^2 & E^3 & \frac{2}{\rho^2}E^1+E^5 & \frac{1}{\rho^2}E^2+\frac{1}{\rho^2}E^3+E^4 \\
\end{bmatrix}
\end{equation}
\begin{equation} \label{eq_NMM_EF}
R_{>2}^{tri}=
2\begin{bmatrix}
\frac{E^1}{2} & E^1 & E^1 & E^1 & E^1 \\
E^1 & E^2 & 2E^1 & E^2 & E^2 \\
E^1 & 2E^1 & E^3 & E^3 & E^3 \\
E^1 & E^2 & E^3 & E^4 & E^5 \\
E^1 & E^2 & E^3 & E^5 & E^4 \\
\end{bmatrix}
\end{equation}

\subsubsection{Analysis of limits and algebras}

\begin{figure}[!htb] 
\begin{center}
\includegraphics[scale=0.5]{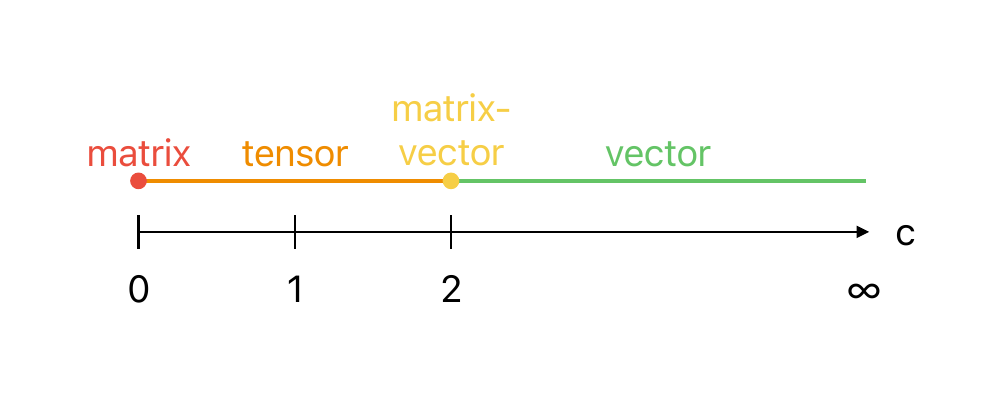}
\caption{The distribution of the algebras on the spectrum of $c$ for the $O(N)^2\times O(M)$ model. Dots represent critical algebras. Lines represent reduced algebras. Recall $M\propto N^c$. The labels show which type of limit yields the algebra.} 
\label{Fig:CspecNNM}
\end{center}
\end{figure}

Below we discuss the properties of each algebra, in the order of how they are distributed on the spectrum of $c$. Figure \ref{Fig:CspecNNM} shows this distribution. 
Most of these four distinct limits have been identified before. 

First of all, the critical large $N$ finite $M$ limit, which yields algebra $C_0^{tri}$ (\ref{eq_NMM_A1})-(\ref{eq_NMM_A2}), corresponds to the matrix-like limit of the trifundamental model \cite{Benedetti:2020sye}. 
In this limit we have two equal large group size-parameters $N_2=N_3=N$, spanning large $N \times N$ matrices, and $N_1=M$ is finite.\footnote{In \cite{Benedetti:2020sye} they consider rectangular matrices while we have quadratic matrices. This does not seem to affect which terms survive the limit to one-loop order.} 
Note that the scaling used in \cite{Benedetti:2020sye} is the same as our scaling (\ref{eq_NMM_scaling1}) with respect to the large group size-parameter $N$. 
Taking the large $M$ limit of this algebra simplifies it into the reduced limit algebra (\ref{eq_NMM_B}).

The reduced limit algebra $R^{tri}_{0<c<2}$ (\ref{eq_NMM_B}) also corresponds to a known limit. It is the result of the large $M$ and $N$ limits where $0<c<2$ for $M=\frac{v(bc)}{w(b)^c}N^c$. Note that, for $c=1$ and $v(b)=w(b)$ we have $M=N$ and therefore the symmetry $O(N)^3$. The $O(N)^3$ model is also called the homogeneous trifundamental model and it has a so called tensor-like large $N$ limit \cite{Benedetti:2020sye}. 
It has been established that the large $N$ limit of the $O(N)^3$ model is dominated by melonic Feynman diagrams and that the tetrahedral coupling $\lambda_5$ is the cause of this dominance since it alone can generate the other couplings, e.g. via RG flow \cite{Bonzom:2011zz,Giombi:2017dtl,Benedetti:2020sye}.

Since this algebra (\ref{eq_NMM_B}) has no free parameter it should be equivalent to the $O(N)^3$ model large $N$ limit algebra for all values $0<c<2$. 
We find that by setting $M=N$ in the scaling (\ref{eq_NMM_scaling1}) it becomes the optimal large $N$ scaling for the $O(N)^3$ model, which was developed in \cite{Carrozza:2015adg} and has been applied in for example \cite{Jepsen:2023pzm,Giombi:2017dtl,Benedetti:2020sye}. 
The algebra $R^{tri}_{0<c<2}$ (\ref{eq_NMM_B}) of the reduced limit is indeed the melonic limit large $N$ algebra of the $O(N)^3$ model when comparing to results in \cite{Benedetti:2020sye} and \cite{Jepsen:2023pzm}. 
Therefore, we have shown that for the $O(N)^2\times O(M)$ model a spectrum of large $N$ and $M$ scalings lead to the melonic limit algebra (i.e. one-loop order beta functions) of the $O(N)^3$ model. The spectrum of scalings are given by the optimal scaling (\ref{eq_NMM_scaling1}) with a large $N$ and $M$ related by $M\propto N^c$ for all $0<c<2$, which shows that at least to one-loop order the melonic limit is robust  since more values than $c=1$ yield the melonic limit. 

Our reduced limit where $0<c<2$ is also comparable to the limit of \cite{Ferrari:2017jgw} with their optimal scaling, because the large $N$ and $M$ limits do not commute and $N^2 \gg M$. In the $O(N)^2 \times O(D)^r$ models with the optimal scaling of \cite{Ferrari:2017jgw} the interpretation of the limit where $N\rightarrow \infty$ followed by $D \rightarrow \infty$ is that the first limit corresponds to the planar limit of matrix models, and the second limit gives a large $D$ expansion for each order in $1/N$ of the planar limit. However, often taking this large $N$ followed by large $D$ limit leads to the melonic limit of the $O(N)^3$ model \cite{Benedetti:2020iyz}. Our results are in agreement with this since we find that our reduced limit leads to the melonic limit algebra.

Next, the critical limit algebra $C^{tri}_{2}$ (\ref{eq_NMM_C}) (or equivalently (\ref{eq_NMM_D})) occurs for large $M$ and $N$ when $M/N^2=\rho^2$ is fixed. 
This appears to be a matrix-vector-like limit since the symmetry $O(N)^2$ describes matrices of the size $N \times N$ while $O(N^2)$ describes a vector of size $N^2$. Note that the algebra of this limit is simpler than the algebra $C^{tri}_{0}$ of the matrix-like limit but more complicated than the vector-like and tensor-like limit algebras $R^{tri}_{>2}$ and $R^{tri}_{0<c<2}$. 
In \cite{Benedetti:2020iyz} 
a model with complex matrices and symmetry $U(N)^2\times O(D)$ is found to have a Veneziano-like limit where $N^2/D$ is fixed, which has a complete recursive characterisation. It is an open question how similar the details of the limits of these models are. 

The final reduced limit algebra $R^{tri}_{c>2}$ (\ref{eq_NMM_EF}) is obtained in the limit where $M \gg N^2$ and corresponds to the vector-like limit of the trifundamental tensor model $O(N_1)\times O(N_2) \times O(N_3)$ considered in \cite{Benedetti:2020sye}. 
It is vector-like since $M \gg N^2$ is effectively vector-like and $M$ grows much faster than $N$, and it includes the finite $N$ large $M$ limit. 
Again, the scaling used in \cite{Benedetti:2020sye} is the same as our scaling (\ref{eq_NMM_scaling2}) with respect to the large group size-parameter. 

There are a few things to note about all these limit algebras. To begin with, the tetrahedron element $E^5$ is rarely generated in the algebras.
For the trifundamental model it can generate all the other couplings by RG flow \cite{Benedetti:2020sye} and we observe this in the (not rescaled) algebra before we take any large $N$ and/or $M$ limits. In the algebras (\ref{eq_NMM_A1})-(\ref{eq_NMM_D}) we still observe this behaviour, but it is modified in the large $N$ and $M$ limits where $M \gg N^2$, i.e. the vector-like limit algebra (\ref{eq_NMM_EF}), where the tetrahedron element $E^5$ only generates the pillow element $E^4$ and neither of the other three elements generates either of the elements $E^4$ or $E^5$. 
In all limit algebras, the element $E^5$ is the only element that does not generate itself directly. In algebraic terms this means that it is always nilpotent in the quotient algebras where the rest of the couplings are modded out, i.e. $E^5 \diamond E^5=0$ for $\mathcal{A}/\{E^1,E^2,E^3,E^4\}$ where $\mathcal{A}$ is the full algebra for any limit. This is in contrast to the other elements which all are idempotent, i.e. generate themselves, in the algebras. See appendix \ref{AppB} for more on quotient algebras. 
 
We also observe that the reduced limit algebras are special limits of the critical limit algebras. That is, for this model the vector-like and tensor-like limits are special limits of the matrix-like and matrix-vector-like limits. Taking $M \rightarrow \infty$ in the matrix-like limit algebra (\ref{eq_NMM_A1})-(\ref{eq_NMM_A2}) or setting $\rho=0$ in the matrix-vector-like algebra (\ref{eq_NMM_C}) both lead to the tensor-like (melonic) limit algebra (\ref{eq_NMM_B}). Similarly, taking $\rho \rightarrow \infty$ in the matrix-vector-like limit algebra (\ref{eq_NMM_D}) gives the vector-like limit algebra (\ref{eq_NMM_EF}).   
 
In summary, we have a matrix-like limit, a tensor-like limit, a matrix-vector-like limit and a vector-like limit. 
The tensor-like limit and the vector-like limit yield simpler algebras than the matrix-like and matrix-vector-like limits, but we note that the matrix-vector-like limit algebra is still slightly simpler than the matrix-like limit algebra. 

\subsection{$SU(N)_A \times O(M)_F$}

In \cite{Flodgren:2023tri} we studied a model with $M$ scalar multiplets in the adjoint $SU(N)$ representation, described by the global symmetry $SU(N)_A\times O(M)_F$. We identified three different limits where $N^2 \geq M$, however, extending the analysis by using the parametrisation (\ref{eq_para}) we are able to identify two additional limits, with $M \gg N^2$.

This model has four invariants\footnote{In \cite{Flodgren:2023lyl,Flodgren:2023tri} we called the basis elements $\{e^{1s},e^{1t},e^{2s},e^{2t}\}$. They are now called $\{e^1,e^2,e^3,e^4\}$ respectively.}, two single-trace ones $\mathcal{O}_{1},\mathcal{O}_{2}$ and two double-trace ones $\mathcal{O}_{3},\mathcal{O}_{4}$. The traces are over the matrix-indices of the $SU(N)$ matrices. The fields can be written $\Phi_{\bar{a}}=\phi_{A}T_{\bar{A}}$ where the multi-indices are $A=\bar{a}\bar{A}$. $\bar{a}=1,\dots,M$ is the scalar multiplet $O(M)$-index and $\bar{A}=1,\dots,N^2-1$ is the $SU(N)$-index. The normalisation of the fundamental representation matrices $T_{\bar{A}}$ is $\Tr(T_{\bar{A}}T_{\bar{B}})=\frac{1}{2}\delta_{\bar{A}\bar{B}}$. 
The invariants and the corresponding symmetric basis elements are related by 
\begin{equation} \label{eq_SUNM_brel}
\begin{aligned}
 \frac{1}{4 !} e_{ABCD}^{1} \phi_{A} \phi_{B} \phi_{C} \phi_{D}&=\frac{1}{2} \Tr \Phi_{\bar{a}} \Phi_{\bar{a}} \Phi_{\bar{b}} \Phi_{\bar{b}} \\
\frac{1}{4 !} e_{ABCD}^{2} \phi_{A} \phi_{B} \phi_{C} \phi_{D}&=\frac{1}{4} \Tr \Phi_{\bar{a}} \Phi_{\bar{b}} \Phi_{\bar{a}} \Phi_{\bar{b}} \\
\frac{1}{4 !} e_{ABCD}^{3} \phi_{A} \phi_{B} \phi_{C} \phi_{D}&=\frac{1}{2} \Tr \Phi_{\bar{a}} \Phi_{\bar{a}} \Tr \Phi_{\bar{b}} \Phi_{\bar{b}}  \\
 \frac{1}{4 !} e_{ABCD}^{4} \phi_{A} \phi_{B} \phi_{C} \phi_{D}&=\Tr \Phi_{\bar{a}} \Phi_{\bar{b}} \Tr \Phi_{\bar{a}} \Phi_{\bar{b}} .
\end{aligned}
\end{equation}
See appendix \ref{AppA2} for the explicit expressions of the basis elements.

\subsubsection{Limits and algebras} \label{sec_SUNM_a}

There are two line segments of optimal scalings parametrised by $c$ that overlap for $c=2$. 
Scaling (\ref{eq_SUNM_scaling1}) is optimal for $0\leq c\leq 2$, i.e. for $N^2 \geq M$,
\begin{align} \label{eq_SUNM_scaling1}
\lambda_1=\frac{\Lambda_1}{MN} && \lambda_2=\frac{\Lambda_2}{\sqrt{M}N} && \lambda_3=\frac{\Lambda_3}{MN^2} && \lambda_4=\frac{\Lambda_4}{N^2},
\end{align}
and scaling (\ref{eq_SUNM_scaling2}) for $c \geq 2$, i.e. for $M \geq N^2$,
\begin{align} \label{eq_SUNM_scaling2}
\lambda_1=\frac{\Lambda_1}{MN} && \lambda_2=\frac{\Lambda_2}{M} && \lambda_3=\frac{\Lambda_3}{MN^2} && \lambda_4=\frac{\Lambda_4}{M}.
\end{align}
Note that $E^1$ and $E^3$ scale identically in both cases, indicating that the behaviour of $E^2$ and $E^4$ changes the most between the limits. For each scaling the large $N$ and $M$ limits do not commute.

There are five large $M$ and/or $N$ limits that yield distinct algebras because the values $c=0$, $c=2$ and $c\rightarrow \infty$ all correspond to critical limits, while the rest of the values of $c$ correspond to reduced limits. The overlap of the scalings at $c=2$ means that the algebras in this limit are equivalent up to a rescaling of the couplings.
The limits are listed below.
\begin{itemize}
\item \textbf{Critical limit $c=0$:} 
a matrix-like large $N$ limit where $M$ is a constant free parameter. The algebra is (\ref{eq_SUNM_1}). Taking the limit $M \rightarrow \infty$ gives the reduced limit algebra with $0<c<2$ (\ref{eq_SUNM_2}). 
\item \textbf{Reduced limit $0<c<2:$} a large $M$ and $N$ limit where $M \propto N^c$ for $0<c<2$, i.e. $N^2 \gg M$. 
The algebra is (\ref{eq_SUNM_2}). The limit is tensor-like in the sense that both $N$ and $M$ are large. 
This limit is possibly similar to the melonic limit, see section \ref{SUNM_analysis}. 
\item \textbf{Critical limit $c=2$:} a matrix-vector-like large $N$ and $M$ limit where $M/N^2=\sigma$ is fixed. $\sigma\geq 0$ is a scaling coefficient defined as $\sigma \equiv \frac{v(2b)}{w(b)^2}$ via the relation (\ref{eq_MNc}) for $c=2$. Scalings (\ref{eq_SUNM_scaling1}) and (\ref{eq_SUNM_scaling2}), respectively yield the equivalent algebras (\ref{eq_SUNM_3}) and (\ref{eq_SUNM_4}) in this limit. Setting $\sigma=0$ in (\ref{eq_SUNM_3}) gives the reduced limit algebra with $0<c<2$ (\ref{eq_SUNM_2}), and taking $\sigma \rightarrow \infty$ in (\ref{eq_SUNM_4}) gives the reduced limit algebra with $c>2$ (\ref{eq_SUNM_5}).
\item \textbf{Reduced limit $2<c<\infty$:} a large $M$ and $N$ limit where $M\propto N^c$ for $c>2$, i.e. $M \gg N^2$. 
The algebra is (\ref{eq_SUNM_5}). The limit is tensor-like in the sense that both $N$ and $M$ are large. 
This limit resembles an enhanced melonic limit, see section \ref{SUNM_analysis}. 
\item \textbf{Critical limit $c\rightarrow \infty$:} a vector-like large $M$ limit where $N$ is a constant free parameter. The algebra is (\ref{eq_SUNM_6}). Taking the limit $N \rightarrow \infty$ gives the reduced algebra with $2<c<\infty$ (\ref{eq_SUNM_5}).
\end{itemize}
The algebras are given by the multiplication tables (\ref{eq_SUNM_1})-(\ref{eq_SUNM_6}), where $adj$ stands for the $SU(N)_A \times O(M)_F$ model, which uses an adjoint representation. 
\begin{equation} \label{eq_SUNM_1}
C_0^{adj}= \begin{bmatrix}
(\frac{1}{2}+\frac{3}{2M})E^1+(\frac{1}{2}+\frac{3}{2M})E^3+\frac{E^4}{2M^2} & \frac{E^1}{2\sqrt{M}}+\frac{E^2}{M}+\frac{E^3}{2\sqrt{M}}+\frac{E^4}{2M^{3/2}} & (1+\frac{1}{M})E^3 & 2E^3+\frac{E^4}{M} \\
\frac{E^1}{2\sqrt{M}}+\frac{E^2}{M}+\frac{E^3}{2\sqrt{M}}+\frac{E^4}{2M^{3/2}} & \frac{E^1}{2}+(\frac{1}{8}+\frac{1}{4M})E^4 & \frac{E^3}{\sqrt{M}} & \frac{E^4}{\sqrt{M}} \\
(1+\frac{1}{M})E^3 & \frac{E^3}{\sqrt{M}} & E^3 & 2E^3 \\
2E^3+\frac{E^4}{M} & \frac{E^4}{\sqrt{M}} & 2E^3 & 2E^4
\end{bmatrix}
\end{equation}
\begin{equation} \label{eq_SUNM_2}
R^{adj}_{0<c<2} = \begin{bmatrix}
\frac{E^1}{2}+\frac{E^3}{2} & 0 & E^3 & 2E^3 \\
0 & \frac{E^1}{2}+\frac{E^4}{8} & 0 & 0 \\
E^3 & 0 & E^3 & 2E^3 \\
2E^3 & 0 & 2E^3 & 2E^4
\end{bmatrix}
\end{equation}
\begin{equation} \label{eq_SUNM_3}
C_{2a}^{adj}= \begin{bmatrix}
\frac{E^1}{2}+\frac{E^3}{2} & 0 & E^3 & 2\sigma E^1+2E^3 \\
0 & (\frac{1}{2}-\frac{\sigma}{2})E^1 +\frac{\sigma}{4}E^3+\frac{E^4}{8} & 0 & 2\sigma E^2 \\
E^3 & 0 & E^3 & (2+2\sigma)E^3 \\
2\sigma E^1+2E^3 & 2\sigma E^2 & (2+2\sigma)E^3 & 12\sigma E^3+(2+2\sigma)E^4
\end{bmatrix}
\end{equation}
\begin{equation} \label{eq_SUNM_4}
C_{2b}^{adj}= \begin{bmatrix}
\frac{E^1}{2}+\frac{E^3}{2} & 0 & E^3 & 2E^1+\frac{2}{\sigma}E^3 \\
0 & (-\frac{1}{2}+\frac{1}{2\sigma})E^1 +\frac{E^3}{4}+\frac{E^4}{8} & 0 & 2E^2 \\
E^3 & 0 & E^3 & (2+\frac{2}{\sigma})E^3 \\
2E^1+\frac{2}{\sigma}E^3 & 2E^2 & (2+\frac{2}{\sigma})E^3 & \frac{12}{\sigma}E^3+(2+\frac{2}{\sigma})E^4
\end{bmatrix}
\end{equation}
\begin{equation} \label{eq_SUNM_5}
R^{adj}_{2<c<\infty}= \begin{bmatrix}
\frac{E^1}{2}+\frac{E^3}{2} & 0 & E^3 & 2E^1 \\
0 & -\frac{E^1}{2}+\frac{E^3}{4}+\frac{E^4}{8} & 0 & 2E^2 \\
E^3 & 0 & E^3 & 2E^3 \\
2E^1 & 2E^2 & 2E^3 & 2E^4
\end{bmatrix}
\end{equation}
\begin{equation} \label{eq_SUNM_6}
C^{adj}_{\infty}= \begin{bmatrix}
(\frac{1}{2}-\frac{2}{N^2})E^1 +(\frac{1}{2}+\frac{1}{N^2})E^3 & -\frac{E^1}{N}+\frac{E^3}{2N} & (1-\frac{1}{N^2})E^3 & 2E^1 \\
-\frac{E^1}{N}+\frac{E^3}{2N} & -\frac{E^1}{2} + \frac{E^3}{4}+\frac{E^4}{8} & -\frac{E^3}{2N} & 2E^2 \\
(1-\frac{1}{N^2})E^3 & -\frac{E^3}{2N} & (1-\frac{1}{N^2})E^3 & 2E^3 \\
2E^1 & 2E^2 & 2E^3 & 2E^4
\end{bmatrix}
\end{equation}

\subsubsection{Analysis of the limits and algebras}\label{SUNM_analysis}

\begin{figure}[!htb] 
\begin{center}
\includegraphics[scale=0.5]{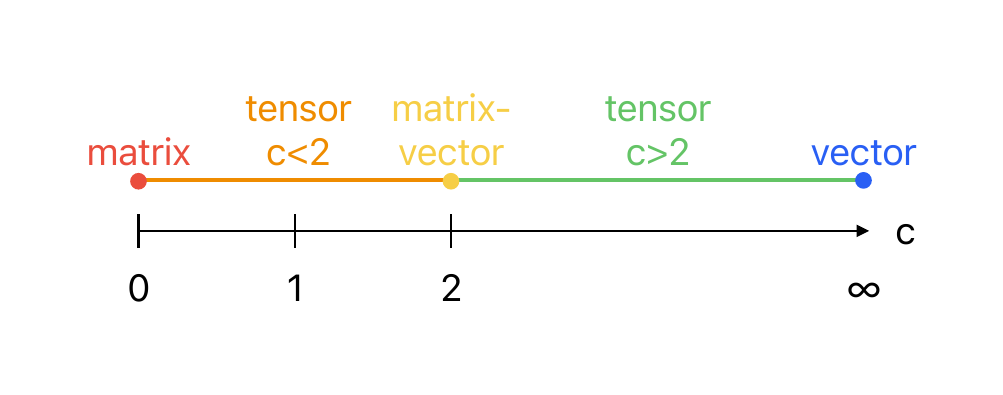}
\caption{The distribution of the algebras on the spectrum of $c$ for the $SU(N)\times O(M)$ model. Dots represent critical algebras. Lines represent reduced algebras. Recall $M\propto N^c$. The labels show which type of limit yields the algebra.}
\label{Fig:CspecSUNM}
\end{center}
\end{figure}

Next we discuss the properties of the algebras, in the order of how they are distributed on the spectrum of $c$ which is shown in figure \ref{Fig:CspecSUNM}.
The limits for which $M \leq N^2$ have been identified and analysed in \cite{Flodgren:2023tri}. 
However, the author has not found the two limits identified via scaling (\ref{eq_SUNM_scaling2}) for which $M \gg N^2$, i.e. algebras $R^{adj}_{2<c<\infty}$ and $C^{adj}_{\infty}$, in the literature for this model. 

First of all, similarly to the previous model, the large $N$ finite $M$ limit, which yields algebra $C^{adj}_{0}$ (\ref{eq_SUNM_1}), is matrix-like in the sense that the $SU(N)$ symmetry corresponds to large $N\times N$ matrices and $M$ is finite. Out of all five algebras this one is the most complicated, but subsequently taking $M \rightarrow \infty$ in the algebra does yield the simpler reduced limit algebra (\ref{eq_SUNM_2}).

The reduced limit algebra $R^{adj}_{0<c<2}$ (\ref{eq_SUNM_2}) shares many properties with the melonic limit algebra (\ref{eq_NMM_B}) of the $O(N)^2\times O(M)$ model.
We observe that in both (\ref{eq_SUNM_2}) and (\ref{eq_NMM_B}) the elements $E^2_{adj}$ and $E^5_{tri}$, for the algebras respectively, 
generate the other elements but are not generated themselves. In addition, the double-trace elements $E^3_{adj}$ in (\ref{eq_SUNM_2}) and $E^1_{tri}$ in (\ref{eq_NMM_B}) share that all off-diagonal products in both algebras only contain these elements. 
These similarities suggest the possibility that for the $SU(N)\times O(M)$ model the large $N$ and $M$ limit where $M\propto N^c$ for $0<c<2$ is similar to the melonic limit in terms of simplicity, i.e. possibly dominated by one type of diagram constructed out of the single-trace interaction with coupling constant $\lambda_2$, even to higher loop orders.

Next, the critical limit that gives the algebra $C^{adj}_{2}$ ((\ref{eq_SUNM_3}) or (\ref{eq_SUNM_4})) is Veneziano-like since both $N$ and $M$ are large while $M/N^2=\sigma$ is fixed, and matrix-vector-like since $SU(N)$ describes large $N\times N$ matrices in the adjoint representation and $O(N^2)$ describes a large vector of size $N^2$. We note that for $\sigma=1$ the algebra loses one term and the symmetry of the model becomes $SU(N)\times O(N^2)$.\footnote{In \cite{Flodgren:2023tri} we noted that $\sigma=1$ shows up as special in the RG flow plots.} 

The second reduced limit algebra $R^{adj}_{2<c<\infty}$ (\ref{eq_SUNM_5}) appears in a large $N$ and $M$ limit where $M \gg N^2$. It is intriguing because it is almost as simple as the other reduced limit algebra $R^{adj}_{0<c<2}$ (\ref{eq_SUNM_2}), which shares properties with the melonic limit algebra of the $O(N)^2\times O(M)$ model. The algebra $R^{adj}_{2<c<\infty}$ appears in a different limit to $R^{adj}_{0<c<2}$ while still being simpler than the algebras in the vector-like, matrix-vector-like and matrix-like limits. 
Note that the two algebras are obtained from different optimal scalings and that they are not special cases of one another since they lack free parameters.  
A noticeable difference between the two 
is that in $R^{adj}_{0<c<2}$ the element $E^2$ is not generated at all but in $R^{adj}_{2<c<\infty}$ the element $E^2$ is generated, 
 which is what suggests that the limit yielding $R^{adj}_{2<c<\infty}$ might be more complicated than the one yielding $R^{adj}_{0<c<2}$. 
We can then speculate that the reduced limit where $M\gg N^2$ could potentially be a sort of ``extended melonic'' limit to the reduced ``melonic'' limit where $N^2 \gg M$.

The last limit is the critical vector-like large $M$ limit where $N$ is finite, which yields the algebra $C^{adj}_{\infty}$ (\ref{eq_SUNM_6}). The free parameter $N$ corresponds to a finite number of $O(M)$ vector models with large $M$. The limit is in this sense vector-like, but with a set of vectors that transform as matrices under $SU(N)$. 
Similar models, so-called Hermitian matrix-vector models with symmetry $U(N)\times O(D)$ have been considered in the large $D$ finite $N$ limit, and have been argued to degenerate into vector models \cite{Carrozza:2020eaz}. However, it is not clear if the scaling used in these cases is the same as the one we identify (\ref{eq_SUNM_scaling2}). 
Using our scaling the vector-like limit seems to yield an algebra of similar complexity to the matrix-vector-like limit algebra of this model, and not a simpler reduced limit algebra like the vector-like limit does in the  $O(N)^2\times O(M)$ model (\ref{eq_NMM_EF}), partly because it contains $N$ as a free parameter.
Taking $N \rightarrow \infty$ in $C^{adj}_{\infty}$ simplifies it to the reduced limit algebra $R^{adj}_{2<c<\infty}$ (\ref{eq_SUNM_5}).

One of the interesting properties of these algebras as a group is how the single-trace element $E^2_{adj}$ behaves. 
We have already noted the similarities between the single-trace element $E^2_{adj}$ for this model and the tetrahedron element $E^5_{tri}$ in the $O(N)^2\times O(M)$ model when it comes to the reduced limit algebra where $N^2 \gg M$, but there are even more similarities. 
The element $E^2_{adj}$ rarely appears in any of the algebras, and is not generated in the reduced limit algebra (\ref{eq_SUNM_2}) at all. 
From the finite $N$ and $M$ algebra we know that the couplings $\lambda_1$ and $\lambda_2$ both generate all four couplings, similarly to the tetrahedron coupling generating all other couplings in the trifundamental model.
However, in the algebras (\ref{eq_SUNM_1})-(\ref{eq_SUNM_6}) $E^2_{adj}$ alone generates all the other elements. 
Furthermore, the element $E^2_{adj}$ is the only element that does not generate itself directly in the algebras, meaning it is always nilpotent in the quotient algebras where the other three couplings are modded out, i.e. $E^2 \diamond E^2=0$ for $\mathcal{A}/\{E^1,E^3,E^4\}$, while the other elements are idempotent in their quotient algebras.  
This is the same as how $E^5_{tri}$ behaves in the $O(N)^2\times O(M)$ model.

To summarise, the three limit algebras of greater complexity are the matrix-like, matrix-vector-like and the vector-like limit algebras, for which we note that the matrix-vector-like one is simpler than the matrix-like one. The two reduced limit algebras are even simpler, with the limit algebra where $N^2 \gg M$ being the simplest and potentially a version of a melonic limit, and the limit algebra where $M \gg N^2$ being slightly more complicated and potentially an enhanced melonic limit. Note that each of the reduced limits can be obtained as special limits of two of the critical limits. 

\subsection{Comparison of models}\label{sec_comparison}

\begin{figure}[!htb] 
\begin{center}
\includegraphics[scale=0.5]{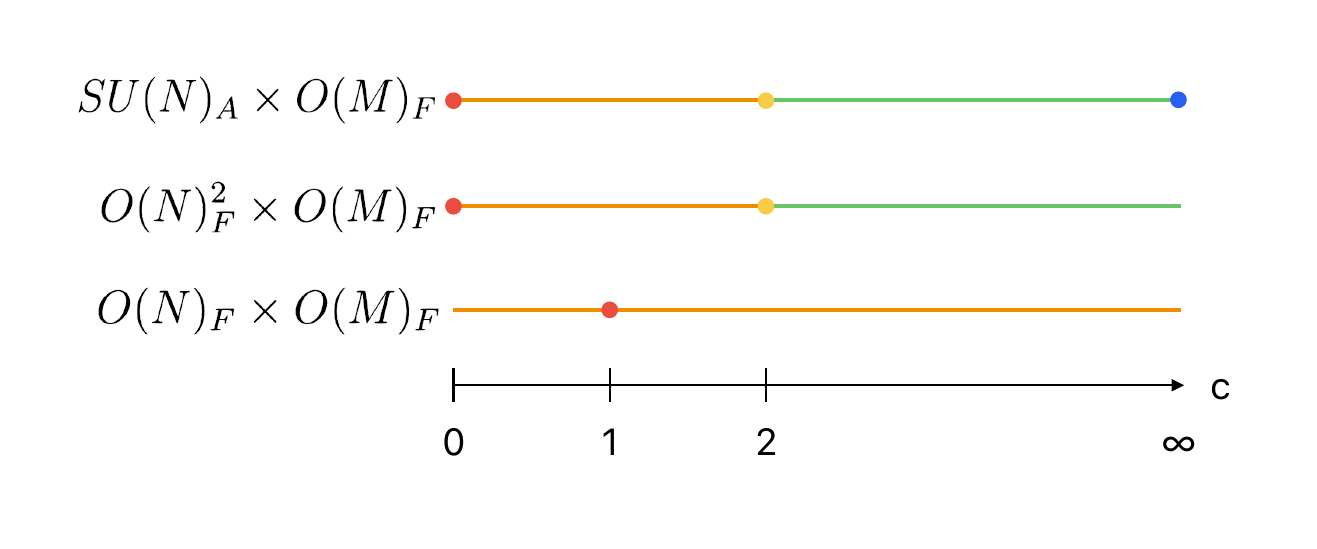}
\caption{The distribution of the algebras on the spectrum of $c$ for the three models. Dots represent critical algebras. Lines represent reduced algebras. For each model the same color represent identical algebras. Recall $M\propto N^c$ and that $c=0$ corresponds to large $N$, finite $M$ and that $c\rightarrow \infty$ corresponds to finite $N$, large $M$.}
\label{Fig:Cspec}
\end{center}
\end{figure}

There are several similarities between the limits of the models we have analysed, including the bifundamental model. Firstly, in all models there are similarities in how the distinct algebras are distributed on the spectrum of $c$, see figure \ref{Fig:Cspec}. 

All models have well-defined Veneziano-like limits, as well as matrix-like and vector-like limits.\footnote{For the bifundamental model the matrix-like limit is a Veneziano-like limit.}  
For each model with two group size-parameters we find that there are two linear segments of optimal scalings parametrised by $c$, both optimal for a range of values of $c$ (recall $M\propto N^c)$ with an overlap at $c=c_*$. 
Varying $c$ within its range for each optimal scaling can take us between different types of limits, e.g. from a matrix-like via a tensor-like to a matrix-vector-like limit. At $c_*$ we need to change between the scalings in addition to varying $c$, to reach the other limits. 
Alternatively, one can go between the algebras of the limits via their free parameters. Critical limit algebras, which in our examples have free parameters in the form of scaling coefficients or finite $N$ or $M$, become reduced limit algebras for extreme limits of these free parameters. 

It is possible for several limits to yield identical algebras, this is the case for reduced limits since they cover ranges of $c$, but also for the bifundamental model where the two reduced limits yield the same algebra, see figure \ref{Fig:Cspec}. 

We note that the $O(N)\times O(M)$ and $O(N)^2 \times O(M)$ models both have algebras given by the vector-like limits where only $M$ is large, and that these algebras are relatively simple and lack free parameters. 
The $SU(N)\times O(M)$ model does have a limit where $M$ is large but its resulting algebra has $N$ as a free parameter, so it is vector-like in the sense that it has a finite number of large $O(M)$ vectors, and its dynamics seems to depend on $N$, but its algebra is not as simple as the vector-like limit algebras of the other models. 

Similarly, the $O(N)^2\times O(M)$ and $SU(N)\times O(M)$ models each have three distinct algebras in the limits where $N^2 \geq M$. 
In both models the limit where $M/N^2$ is fixed appears as a critical limit. In \cite{Benedetti:2020iyz} a complex multi-matrix\footnote{In our present terminology, this is a matrix-vector model.} model with symmetry $U(N)^2\times O(D)$ is found to have a limit where $N^2/D$ is fixed, where a particular set of diagrams survive.
Therefore, the Veneziano-like limit where $N^2/M$ (or $N^2/D$) is fixed appears in all of these three similar models.

As discussed in section \ref{sec_SUNM_a} both $O(N)^2\times O(M)$ and $SU(N)\times O(M)$ have a coupling that behaves similarly in that it generates the other couplings but rarely appears in the algebras, and in one limit algebra it does not appear at all. For the $O(N)^2\times O(M)$ model this is the tetrahedron coupling, which is the key to the melonic limit. For $SU(N)\times O(M)$ the single-trace coupling\footnote{Which also has a tetrahedron shape in terms of the colour and flavour lines.} $\lambda_2$ has this behaviour to leading order.

\section{Discussion}
We have found that the algebraic description of the one-loop RG flow for multiscalar theories with quartic interactions in four dimensions can be used to identify useful scalings for large $N$ limits. 

For the matrix and tensor models with $O(N)_F^2$ and $O(N)_F^3$ symmetry we have shown that the scaling identified by our method is the optimal scaling identified by others via different methods. 
In the case of models with two group size-parameters, such as the bifundamental model with symmetry $O(N)_F\times O(M)_F$, our method identifies two scalings, each of which is optimal for range of values of $c$, where $c$ signifies the proportionality between $M$ and $N$ via $M\propto N^c$. 
The parametrisation in terms of $c$ allows us to categorise the limits on a spectrum of $c$. 
In this way, we characterise vector-like, matrix-like, tensor-like and Veneziano-like limits of several matrix and tensor models, resulting in a few limits that to the authors knowledge have not been described in the literature. 

One model we consider is the trifundamental tensor model, but with the less general symmetry $O(N)^2_F\times O(M)_F$ compared to $O(N_1)\times O(N_2) \times O(N_3)$. We identify four limits that yield distinct algebras, they are three limits which have been studied in detail in \cite{Benedetti:2020sye}, and one Veneziano-like limit where $M/N^2=\rho^2$ is fixed. The known limits are a vector-like limit, a matrix-like limit and a tensor-like limit that includes the melonic limit. The Veneziano-like limit has the symmetry $O(N)^2_F\times O(\rho^2 N^2)_F$ and we call it matrix-vector-like. A similar Veneziano-like limit was identified for complex matrices and given a complete recursive characterisation in \cite{Benedetti:2020iyz}. The tensor-like limit has the simplest algebra, followed by the vector-like limit, compared to the matrix-vector-like and matrix-like limits. We note that the matrix-vector-like limit yields a slightly simpler algebra than the matrix-like limit. 

For a model with $M$ fields in the adjoint $SU(N)$ representation, i.e. the global symmetry $SU(N)_A\times O(M)_F$, which we previously identified three limits for \cite{Flodgren:2023tri}, we identify two additional limits that yield distinct algebras. One of which is the vector-like  limit with large $M$ and finite $N$. Interestingly, with our scaling this limit does not seem to degenerate the model to a vector model, which is often the case in vector-like limits \cite{Carrozza:2020eaz}. This suggests the possibility that the limit is not dominated by bubble diagrams.  
The other additional limit is a Veneziano-like limit where $M \gg N^2$ but both $N$ and $M$ are large. 
Out of the five distinct algebras there are two that are simpler, similarly to how the melonic limit is relatively simple in tensor models. These are limits with large $N$ and $M$, one where $N^2 \gg M$ and one where $M \gg N^2$, that are both less complicated than the matrix-like, matrix-vector-like and vector-like limits of the same model. The limit with $N^2 \gg M$, yields an algebra with properties similar to the melonic limit of tensor models, in that one of the couplings that can generate the others is itself not generated in the limit, similarly to how the tetrahedron coupling in tensor models behaves. The limit with $M \gg N^2$ yields a slightly more complicated algebra in that all couplings appear in the algebra, but it is still less complicated than the algebras of the remaining three limits, which makes us speculate that it could be an enhanced melonic limit. 

None of the new limits we identified have been studied to higher loop orders by us or elsewhere in the literature we have found. Therefore, it is possible that these scalings and limits do not yield well behaved RG flow equations to higher than one-loop order. However, considering that the same method that yields these new optimal scalings and limits also identifies known optimal scalings and limits, we are hopeful that the new limits are well behaved to at least some higher loop orders as well.

When it comes to applying our method to quartic multiscalar models with different symmetries in 4d, it should be fairly straightforward. For example, to apply our method to the full trifundamental model with symmetry $O(N_1)\times O(N_2) \times O(N_3)$ one needs to make a general ansatz for the relationships between $N_1, N_2$ and $N_3$ to identify any remaining new limits. The main difficulty will be in analysing the larger space of parameters relating $N_1$, $N_2$ and $N_3$, compared to just one parameter relating $N$ and $M$ in the models we have considered.

There are a couple of noteworthy aspects to our method and results. Using the algebra the method leads to a system of inequalities of the scaling exponents, which in practice forms a linear programming problem. Crucially, the redundant conditions of this system are what determines which terms survive in the limits. This fact enables us to not only identify the optimal scaling, but also potentially useful non-optimal scalings. 
In addition, we considered only models with two group size-parameters and for similar models a non-commutativity of limits has been observed \cite{Ferrari:2017ryl,Azeyanagi:2017mre,Ferrari:2017jgw,Benedetti:2020iyz}. Similarly, for the models we consider (with parameters $N$ and $M$) we find that the large $N$ and $M$ limits do not commute and we obtain an explanation of the non-commutativity, in that the inequalities which have to be respected depends on which parameter is taken to be large.

The main benefit of our method is that it requires no diagrammatic or combinatorial calculations. 
However, so far our method only uses one-loop order results. Since it does not guarantee that the optimal scalings it yields would lead to finite large $N$ beta functions at higher loop orders its usefulness therefore comes from requiring less complicated calculations, compared to combinatorial and graph theoretical methods. Potentially interesting limits may be identified quickly, for further scrutiny by other methods.

In future work, we hope to investigate how to apply the algebraic description to higher loop orders. This in turn, could help clarify if the one-loop order method for obtaining the scalings works for higher loop orders, or if it is possible to identify the properties of dominant diagrams in a limit from the algebra. 
There may be additional algebraic concepts and structures that are needed to describe higher loop behaviours of the RG flow and system over all. Possibly ones related to the description of beta functions as a gradient flow \cite{Pannell:2024sia} via a metric. 

We originally developed the algebraic description for marginal $\phi^4$ theory in 4d. A natural future direction is describing theories with different types of interactions than simply quartic scalar ones via algebras, if that is possible.

\appendix

\section{Basis elements and invariants} \label{AppA}
\subsection{$O(N)^2_F\times O(M)_F$}\label{AppA1}

The five quartic invariant polynomials are
\begin{equation} \label{eq_ONNM_invariants}
\begin{split}
\mathcal{O}_1&=(\phi_{a b c} \phi_{a b c})^2 \\
\mathcal{O}_2&=\phi_{a_1 b_1 c_1} \phi_{a_1 b_1 c_2} \phi_{a_2 b_2 c_1} \phi_{a_2 b_2 c_2} \\
\mathcal{O}_3&=\phi_{a_1 b_1 c_1} \phi_{a_1 b_2 c_1} \phi_{a_2 b_1 c_2} \phi_{a_2 b_2 c_2} \\
\mathcal{O}_4&=\phi_{a_1 b_1 c_1} \phi_{a_2 b_1 c_1} \phi_{a_1 b_2 c_2} \phi_{a_2 b_2 c_2} \\
\mathcal{O}_5&=\phi_{a_1 b_1 c_1} \phi_{a_1 b_2 c_2} \phi_{a_2 b_1 c_2} \phi_{a_2 b_2 c_1}.
\end{split}
\end{equation}
The corresponding symmetric basis elements are 
\begin{equation}\label{eq_ONNM_basis}
\begin{split}
e^1_{ABCD} &= \delta_{a_1a_2}\delta_{b_1b_2}\delta_{c_1c_2}\delta_{a_3a_4}\delta_{b_3b_4}\delta_{c_3c_4} +
\delta_{a_1a_3}\delta_{b_1b_3}\delta_{c_1c_3}\delta_{a_2a_4}\delta_{b_2b_4}\delta_{c_2c_4}
+ \delta_{a_1a_4}\delta_{b_1b_4}\delta_{c_1c_4}\delta_{a_2a_3}\delta_{b_2b_3}\delta_{c_2c_3}\\
e^2_{ABCD} &= \delta_{a_1a_2}\delta_{b_1b_2}\delta_{c_1c_3}\delta_{a_3a_4}\delta_{b_3b_4}\delta_{c_2c_4}+ \delta_{a_1a_2}\delta_{b_1b_2}\delta_{c_1c_4}\delta_{a_3a_4}\delta_{b_3b_4}\delta_{c_2c_3} \\
&+ \delta_{a_1a_3}\delta_{b_1b_3}\delta_{c_1c_2}\delta_{a_2a_4}\delta_{b_2b_4}\delta_{c_3c_4} 
+\delta_{a_1a_3}\delta_{b_1b_3}\delta_{c_1c_4}\delta_{a_2a_4}\delta_{b_2b_4}\delta_{c_2c_3} \\
&+ \delta_{a_1a_4}\delta_{b_1b_4}\delta_{c_1c_2}\delta_{a_2a_3}\delta_{b_2b_3}\delta_{c_3c_4} 
+  \delta_{a_1a_4}\delta_{b_1b_4}\delta_{c_1c_3}\delta_{a_2a_3}\delta_{b_2b_3}\delta_{c_2c_4}  \\
e^3_{ABCD} &= \delta_{a_1a_2}\delta_{b_1b_3}\delta_{c_1c_2}\delta_{a_3a_4}\delta_{b_2b_4}\delta_{c_3c_4}
+ \delta_{a_1a_2}\delta_{b_1b_4}\delta_{c_1c_2}\delta_{a_3a_4}\delta_{b_2b_3}\delta_{c_3c_4} \\
&+ \delta_{a_1a_3}\delta_{b_1b_2}\delta_{c_1c_3}\delta_{a_2a_4}\delta_{b_3b_4}\delta_{c_2c_4} 
+ \delta_{a_1a_3}\delta_{b_1b_4}\delta_{c_1c_3}\delta_{a_2a_4}\delta_{b_2b_3}\delta_{c_2c_4} \\
&+ \delta_{a_1a_4}\delta_{b_1b_2}\delta_{c_1c_4}\delta_{a_2a_3}\delta_{b_3b_4}\delta_{c_2c_3} 
+ \delta_{a_1a_4}\delta_{b_1b_3}\delta_{c_1c_4}\delta_{a_2a_3}\delta_{b_2b_4}\delta_{c_2c_3} \\
e^4_{ABCD} &= \delta_{a_1a_3}\delta_{b_1b_2}\delta_{c_1c_2}\delta_{a_2a_4}\delta_{b_3b_4}\delta_{c_3c_4} 
+\delta_{a_1a_4}\delta_{b_1b_2}\delta_{c_1c_2}\delta_{a_2a_3}\delta_{b_3b_4}\delta_{c_3c_4}  \\
&+ \delta_{a_1a_2}\delta_{b_1b_3}\delta_{c_1c_3}\delta_{a_3a_4}\delta_{b_2b_4}\delta_{c_2c_4} 
+ \delta_{a_1a_4}\delta_{b_1b_3}\delta_{c_1c_3}\delta_{a_2a_3}\delta_{b_2b_4}\delta_{c_2c_4} \\
&+ \delta_{a_1a_2}\delta_{b_1b_4}\delta_{c_1c_4}\delta_{a_3a_4}\delta_{b_2b_3}\delta_{c_2c_3} 
+ \delta_{a_1a_3}\delta_{b_1b_4}\delta_{c_1c_4}\delta_{a_2a_4}\delta_{b_2b_3}\delta_{c_2c_3} \\
e^5_{ABCD} &= \delta_{a_1a_2}\delta_{b_1b_3}\delta_{c_1c_4}\delta_{a_3a_4}\delta_{b_2b_4}\delta_{c_2c_3} 
+ \delta_{a_1a_2}\delta_{b_1b_4}\delta_{c_1c_3}\delta_{a_3a_4}\delta_{b_2b_3}\delta_{c_2c_4}  \\
&+ \delta_{a_1a_3}\delta_{b_1b_2}\delta_{c_1c_4}\delta_{a_2a_4}\delta_{b_3b_4}\delta_{c_2c_3} 
+\delta_{a_1a_3}\delta_{b_1b_4}\delta_{c_1c_2}\delta_{a_2a_4}\delta_{b_2b_3}\delta_{c_3c_4}  \\
&+\delta_{a_1a_4}\delta_{b_1b_2}\delta_{c_1c_3}\delta_{a_2a_3}\delta_{b_3b_4}\delta_{c_2c_4} 
+\delta_{a_1a_4}\delta_{b_1b_3}\delta_{c_1c_2}\delta_{a_2a_3}\delta_{b_2b_4}\delta_{c_3c_4}. 
\end{split}
\end{equation}
The relation between the invariants and basis elements is shown in (\ref{eq_ONNM_brel}).

\subsection{$SU(N)_A \times O(M)_F$} \label{AppA2}
The four quartic invariants polynomials are
\begin{equation} \label{eq_polynomials}
\begin{split}
&\Tr \Phi_{\bar{a}} \Phi_{\bar{a}} \Phi_{\bar{b}} \Phi_{\bar{b}} \\
&\Tr \Phi_{\bar{a}} \Phi_{\bar{b}} \Phi_{\bar{a}} \Phi_{\bar{b}} \\ 
&\Tr \Phi_{\bar{a}} \Phi_{\bar{a}} \Tr \Phi_{\bar{b}} \Phi_{\bar{b}} \\ 
&\Tr \Phi_{\bar{a}} \Phi_{\bar{b}} \Tr \Phi_{\bar{a}} \Phi_{\bar{b}}. 
\end{split}
\end{equation}
Their corresponding symmetric basis elements are 
\begin{equation} \label{eq_SUNM_basis}
\begin{aligned}
e^{1}_{ABCD} \equiv & \delta_{\bar{a} \bar{b}} \delta_{\bar{c} \bar{d}}\left(\Tr\left(T_{\bar{A}} T_{\bar{B}} T_{\bar{C}} T_{\bar{D}}\right)+\Tr\left(T_{\bar{A}} T_{\bar{D}} T_{\bar{C}} T_{\bar{B}}\right)+\Tr\left(T_{\bar{A}} T_{\bar{B}} T_{\bar{D}} T_{\bar{C}}\right)+\Tr\left(T_{\bar{A}} T_{\bar{C}} T_{\bar{D}} T_{\bar{B}}\right)\right) \\
& +\delta_{\bar{a} \bar{c}} \delta_{\bar{b} \bar{d}}\left(\Tr\left(T_{\bar{A}} T_{\bar{B}} T_{\bar{D}} T_{\bar{C}}\right)+\Tr\left(T_{\bar{A}} T_{\bar{C}} T_{\bar{D}} T_{\bar{B}}\right)+\Tr\left(T_{\bar{A}} T_{\bar{C}} T_{\bar{B}} T_{\bar{D}}\right)+\Tr\left(T_{\bar{A}} T_{\bar{D}} T_{\bar{B}} T_{\bar{C}}\right)\right) \\
& +\delta_{\bar{a} \bar{d}} \delta_{\bar{b} \bar{c}}\left(\Tr\left(T_{\bar{A}} T_{\bar{C}} T_{\bar{B}} T_{\bar{D}}\right)+\Tr\left(T_{\bar{A}} T_{\bar{D}} T_{\bar{B}} T_{\bar{C}}\right)+\Tr\left(T_{\bar{A}} T_{\bar{B}} T_{\bar{C}} T_{\bar{D}}\right)+\Tr\left(T_{\bar{A}} T_{\bar{D}} T_{\bar{C}} T_{\bar{B}}\right)\right) \\
e^{2}_{ABCD} \equiv & \delta_{\bar{a} \bar{b}} \delta_{\bar{c} \bar{d}}\left(\Tr\left(T_{\bar{A}} T_{\bar{C}} T_{\bar{B}} T_{\bar{D}}\right)+\Tr\left(T_{\bar{A}} T_{\bar{D}} T_{\bar{B}} T_{\bar{C}}\right)\right)+\delta_{\bar{a} \bar{c}} \delta_{\bar{b} \bar{d}}\left(\Tr\left(T_{\bar{A}} T_{\bar{B}} T_{\bar{C}} T_{\bar{D}}\right)+\Tr\left(T_{\bar{A}} T_{\bar{D}} T_{\bar{C}} T_{\bar{B}}\right)\right) \\
& +\delta_{\bar{a} \bar{d}} \delta_{\bar{b} \bar{c}}\left(\Tr\left(T_{\bar{A}} T_{\bar{B}} T_{\bar{D}} T_{\bar{C}}\right)+\Tr\left(T_{\bar{A}} T_{\bar{C}} T_{\bar{D}} T_{\bar{B}}\right)\right) \\
e^{3}_{ABCD} \equiv & \delta_{\bar{a} \bar{b}} \delta_{\bar{c} \bar{d}} \delta_{\bar{A} \bar{B}} \delta_{\bar{C} \bar{D}}+\delta_{\bar{a} \bar{c}} \delta_{\bar{b} \bar{d}} \delta_{\bar{A} \bar{C}} \delta_{\bar{B} \bar{D}}+\delta_{\bar{a} \bar{d}} \delta_{\bar{b} \bar{c}} \delta_{\bar{A} \bar{D}} \delta_{\bar{B} \bar{C}} \\
e^{4}_{ABCD} \equiv & \delta_{\bar{a} \bar{b}} \delta_{\bar{c} \bar{d}}\left(\delta_{\bar{A} \bar{C}} \delta_{\bar{B} \bar{D}}+\delta_{\bar{A} \bar{D}} \delta_{\bar{B} \bar{C}}\right)+\delta_{\bar{a} \bar{c}} \delta_{\bar{b} \bar{d}}\left(\delta_{\bar{A} \bar{B}} \delta_{\bar{C} \bar{D}}+\delta_{\bar{A} \bar{D}} \delta_{\bar{C} \bar{B}}\right)+\delta_{\bar{a} \bar{d}} \delta_{\bar{b} \bar{c}}\left(\delta_{\bar{A} \bar{B}} \delta_{\bar{D} \bar{C}}+\delta_{\bar{A} \bar{C}} \delta_{\bar{D} \bar{B}}\right).
\end{aligned}
\end{equation}
The relation between the invariants and basis elements is shown in (\ref{eq_SUNM_brel}).

\section{Subalgebras and ideals}\label{AppB}
This is a very brief review of the subalgebras and ideals of an algebra. A more detailed discussion is found in \cite{Flodgren:2023tri}. 
We denote a linear space spanned by elements with the brackets by $\{ \dots \}$. The full algebra for any model and limit is denoted $\mathcal{A}= \{ e^1, \dots \}$. 

Subalgebras are linear subspaces of the full algebra that form a closed algebra under multiplication with the $\diamond$ product, i.e. the elements of a subalgebra only induce each other.  

Ideals are subalgebras with the requirement that a product of any element of the full algebra with an element in the ideal belongs to the ideal. Via ideals we define quotient algebras, which are obtained by modding out an ideal of the full algebra, e.g. a quotient algebra of the ideal $\{e^1\}$ is $\mathcal{A}/\{e^1\}$ and does not contain the element $e^1$ at all. The RG equations for the couplings corresponding to the elements of a quotient algebra form a closed dynamical system, independent of the couplings of the ideal. 

The subalgebras and ideals we list here are only the ones made up of sets of individual basis elements, i.e. we do not write down the subalgebras and ideals made of linear combinations of the basis elements. In \cite{Flodgren:2023tri} we called these symmetry-respecting subspaces and defined them as subspaces spanned by one or several basis elements of the symmetry-respecting basis (which is the kind of basis with symmetric basis elements, which we use here). If one considers all linear combinations of basis elements there are potentially more subalgebras and ideals.

In previous work we have observed that the finite algebra (before any limits are taken) has subalgebras but not ideals. The ideals appear in the limits of large $N$ and/or $M$. This is still the case for the models we considered here. 

\subsection{$O(N)_F\times O(M)_F$} \label{AppB0}

The finite $N$ and $M$ algebra (\ref{eq_NM_alg1}) has one subalgebra $\{e^1\}$ and no ideals. 
The large $N$ and/or $M$ limit algebras listed in section \ref{sec_ONM_final} each have subalgebras and ideals shown in table \ref{TONM_a}. 
 \begin{table}[h]
\centering
\begin{tabular}{|c|c|c|}
\hline
Algebra & Subalgebras & Ideals \\
\hline
Reduced (\ref{eq_ONM_mat1}) & $\{E^1\}$ $\{E^2\}$ & $\{E^1\}$ \\
\hline
Critical (\ref{eq_ONM_mat2}) & $\{E^1\}$ & $\{E^1\}$ \\
\hline
\end{tabular}
\caption{Algebra properties of the $O(N)_F\times O(M)_F$ model.}
\label{TONM_a}
\end{table}

\subsection{$O(N)_F^2\times O(M)_F$} \label{AppB1}

The finite $N$ and $M$ algebra has the subalgebras $\{e^1\}$, $\{e^1,e^2\}$, $\{e^1,e^3\}$, $\{e^1,e^4\}$ and no ideals. 

The large $N$ and/or $M$ limit algebras listed in section \ref{sec_ONMM_a} each have subalgebras and ideals shown in table \ref{TONNM_a}. 
 \begin{table}[h]
\centering
\begin{tabular}{|c|m{6cm}|m{5cm}|}
  \hline
Algebra & Subalgebras & Ideals \\
\hline
$C^{tri}_{0}$ (\ref{eq_NMM_A1})-(\ref{eq_NMM_A2}) & $\{E^1\}$ $\{E^4\}$ $\{E^1,E^2\}$ $\{E^1,E^3\}$ $\{E^1,E^4\}$ $\{E^1,E^2,E^4\}$ $\{E^1,E^3, E^4\}$ $\{E^1,E^2,E^3, E^4\}$ & $\{E^1\}$ $\{E^1,E^4\}$ \\
\hline
$R^{tri}_{0<c<2}$ (\ref{eq_NMM_B}) & $\{E^1\}$ $\{E^2\}$ $\{E^3\}$ $\{E^4\}$ $\{E^1,E^2\}$ $\{E^1,E^3\}$ $\{E^1, E^4\}$ $\{E^1,E^2,E^3\}$ $\{E^1,E^2,E^4\}$ $\{E^1, E^3,E^4\}$ $\{E^1,E^2,E^3,E^4\}$  & $\{E^1\}$ $\{E^1,E^2\}$ $\{E^1,E^3\}$ $\{E^1,E^4\}$  $\{E^1,E^2,E^3\}$ $\{E^1,E^2,E^4\}$ $\{E^1, E^3,E^4\}$ $\{E^1,E^2,E^3,E^4\}$ \\
\hline
$C^{tri}_{2}$ (\ref{eq_NMM_C}), (\ref{eq_NMM_D}) & $\{E^1\}$ $\{E^2\}$ $\{E^3\}$ $\{E^1, E^2\}$ $\{E^1,E^3\}$ $\{E^1,E^4\}$ $\{E^1,E^2,E^3\}$ $\{E^1,E^2,E^4\}$ $\{E^1,E^2,E^3,E^4\}$ & $\{E^1\}$ $\{E^1,E^2\}$ $\{E^1,E^3\}$ $\{E^1,E^2,E^3\}$ \\
\hline
$R^{tri}_{>2}$ (\ref{eq_NMM_EF}) & $\{E^1\}$ $\{E^2\}$ $\{E^3\}$ $\{E^4\}$ $\{E^1,E^2\}$ $\{E^1, E^3\}$ $\{E^1,E^4\}$ $\{E^2,E^4\}$ $\{E^3,E^4\}$ $\{E^4,E^5\}$ $\{E^1,E^2, E^3\}$ $\{E^1,E^2,E^4\}$ $\{E^1,E^3, E^4\}$ $\{E^1,E^4,E^5\}$ $\{E^2,E^4,E^5\}$ $\{E^3,E^4,E^5\}$ $\{E^1,E^2, E^3, E^4\}$ $\{E^1,E^2,E^4,E^5\}$ $\{E^1,E^3,E^4,E^5\}$ & $\{E^1\}$ $\{E^1,E^2\}$ $\{E^1, E^3\}$  $\{E^1,E^2, E^3\}$ \\
\hline
\end{tabular}
\caption{Algebra properties of the $O(N)_F^2\times O(M)_F$ model.} 
\label{TONNM_a}
\end{table}
We note that the limits where $M \geq N^2$ yield algebras with the same set of symmetry-respecting ideals, but different sets of subalgebras.

\subsection{$SU(N)_A\times O(M)_F$}\label{AppB2}

The finite $N$ and $M$ algebra has the subalgebras $\{e^3\}$, $\{e^3,e^4\}$ and no ideals. 

The large $N$ and/or $M$ limit algebras listed in section \ref{sec_SUNM_a} each have subalgebras and ideals shown in table \ref{TSUNM_a}.
\begin{table}[htbp]
\centering
\begin{tabular}{|c|m{6cm}|m{5cm}|}
\hline
Algebra & Subalgebras & Ideals \\
\hline
\hline
$C^{adj}_{0}$ (\ref{eq_SUNM_1}) & $\{E^3\}$ $\{E^4\}$ $\{E^3,E^4\}$ $\{E^1,E^3,E^4\}$ & $\{E^3\}$ $\{E^3,E^4\}$ \\
\hline
$R^{adj}_{0<c<2}$ (\ref{eq_SUNM_2}) & $\{E^3\}$ $\{E^4\}$ $\{E^1,E^3\}$ $\{E^3,E^4\}$ $\{E^1,E^3,E^4\}$ & $\{E^3\}$ $\{E^1,E^3\}$ $\{E^3,E^4\}$ $\{E^1,E^3,E^4\}$ \\
\hline
$C^{adj}_{2}$ (\ref{eq_SUNM_3}), (\ref{eq_SUNM_4}) & $\{E^3\}$ $\{E^1,E^3\}$ $\{E^3,E^4\}$ $\{E^1,E^3,E^4\}$ & $\{E^3\}$ $\{E^1,E^3\}$ \\
\hline
$R^{adj}_{2<c<\infty}$ (\ref{eq_SUNM_5}) & $\{E^3\}$ $\{E^4\}$ $\{E^1,E^3\}$ $\{E^3,E^4\}$ $\{E^1,E^3,E^4\}$ & $\{E^3\}$ $\{E^1,E^3\}$ \\
\hline
$C^{adj}_{\infty}$ (\ref{eq_SUNM_6}) & $\{E^3\}$ $\{E^4\}$ $\{E^1, E^3\}$ $\{E^3,E^4\}$ $\{E^1, E^3, E^4\}$ & $\{E^3\}$ $\{E^1,E^3\}$ \\
\hline
\end{tabular}
\caption{Algebra properties of the $SU(N)_A\times O(M)_F$ model.}
\label{TSUNM_a}
\end{table}
We note that the limits where $M \geq N^2$ yield algebras with the same set of symmetry-respecting ideals. 
The algebras $R^{adj}_{2<c<\infty}$ and $C^{adj}_{\infty}$ also share the same set of symmetry-respecting subalgebras. The differences in the two algebras appear in the $E^1 \diamond E^2$ and $E^2 \diamond E^3$ interactions, but these do not imply any differences in the symmetry-respecting subalgebras and ideals.

\section{Non-optimal scalings}\label{AppC}

For the bifundamental model we noted that there are non-optimal solutions to the system of exponent conditions (\ref{eq_ineq1}). Technically, the entire set of feasible solutions, which forms a polytope, to an LP problem except the one optimal solution are non-optimal. However, the solutions on the edges of the polytope are most interesting since they correspond to non-trivial algebras, unlike the interior of the polytope. In particular, the solutions at the vertices correspond to more terms surviving the algebra in the limit than do solutions on the edges between vertices. In the case of the bifundamendal model there are two vertex-solutions, one of which is the optimal one. Here we will analyse the non-optimal vertex-solution. 

Table \ref{TONM_exp1} shows which of the exponent conditions are fulfilled by equalities for each range of $c$ for the optimal solution to the system (\ref{eq_ineq1}). Table \ref{TONM_exp2} also includes the non-optimal vertex-solution.
\begin{table}[h]
\centering
\begin{tabular}{|c|c|c|c|c|c|}
  \hline
Exp. cond. & $0 \leq c<1 \text{ (optimal)}$ & $0 \leq c<1$ & $c=1$ & $c>1 \text{ (optimal)}$ & $c>1$  \\ \hline\hline
$\tilde{u}(1)\geq1+c$  & = & > &  =  & = & > \\ \hline
$\tilde{u}(2)\geq 1$  & = & = & =   & >  & > \\ \hline
$\tilde{u}(2)\geq c$  & > & > & =  & = & = \\ \hline
$\tilde{u}(2)\geq \frac{\tilde{u}(1)}{2}$  & > & = & = & > & = \\ \hline
\end{tabular}
\caption{Exponent conditions fulfilled by an equality or strict inequality for the different vertex-solutions of the system of exponent conditions (\ref{eq_ineq1}) for the $O(N)_F\times O(M)_F$ model. Recall $M \propto N^c$.}
\label{TONM_exp2}
\end{table}

In table \ref{TONM_exp2} we note that the same number of conditions are fulfilled by equalities for the two solutions with $0 \leq c <1$ and for the two solutions with $c>1$. This does not necessarily mean that the same number of terms survive the large $Q$ limit as the exponent conditions can appear a different number of times.

Columns 2 and 5 in table \ref{TONM_exp2} are what corresponds to the optimal scalings (\ref{eq_NM_scaling1}) and (\ref{eq_NM_scaling2}). 
The alternative (non-optimal) scalings, obtained from columns 3 and 6 in table \ref{TONM_exp2}, are (\ref{eq_nop1}) and (\ref{eq_nop2}). 
For $0 \leq c \leq 1$, i.e. $N \geq M$, the non-optimal solution is $\tilde{u}(1)=2$, $\tilde{u}(2)=1$, which corresponds to
\begin{align} \label{eq_nop1}
\lambda_1 = \frac{\Lambda_1}{N^2} && \lambda_2 = \frac{\Lambda_2}{N}.
\end{align}
For $c \geq 1$, i.e. $M \geq N$, the non-optimal solution is $\tilde{u}(1)=2c$, $\tilde{u}(2)=c$, which corresponds to
\begin{align} \label{eq_nop2}
\lambda_1 = \frac{\Lambda_1}{M^2} && \lambda_2 = \frac{\Lambda_2}{M}.
\end{align}

For $c=1$ we have a critical limit with $M\propto N$ and all four scalings (\ref{eq_NM_scaling1}), (\ref{eq_NM_scaling2}), (\ref{eq_nop1}) and (\ref{eq_nop2}) should yield equivalent algebras in the limit. 
Using the non-optimal scaling (\ref{eq_nop1}) at  $c=1$ and taking the limit we obtain the algebra 
\begin{equation}\label{eq_ONM_nop1_A}
\begin{bmatrix}
\omega E^1 & 2(1+\omega)E^1 \\
2(1+\omega)E^1 & 12E^1+2(1+\omega)E^2 \\
\end{bmatrix},
\end{equation}
where $\omega \equiv \frac{v(b)}{w(b)}\geq 0$ from (\ref{eq_MNc}) for $c=1$. With the non-optimal scaling (\ref{eq_nop2}) for $c=1$ the limit yields an equivalent algebra that is obtained from (\ref{eq_ONM_nop1_A}) by replacing $\omega \rightarrow \frac{1}{\omega}$. Both are equivalent to (\ref{eq_ONM_mat2}) and are non-associative.

The rest of the limits are reduced limits. Using the non-optimal scalings, i.e. $0 \leq c <1$ and $c>1$ for (\ref{eq_nop1}) and (\ref{eq_nop2}) respectively, yields the algebra
\begin{equation}\label{eq_ONM_nop1_B}
\begin{bmatrix}
0 & 2E^1 \\
2E^1 & 12E^1+2E^2 \\
\end{bmatrix},
\end{equation}
which can also be obtained from (\ref{eq_ONM_nop1_A}) by setting $\omega =0$. Note that this algebra is not equivalent or identical to the reduced limit algebra using the optimal scalings (\ref{eq_ONM_mat1}), although both are associative.

\acknowledgments
The author wishes to thank Bo Sundborg for helpful discussions and feedback. 

\bibliography{BibliographyManuscript}{}
\bibliographystyle{JHEP}

\end{document}